\documentclass{appolb}
\usepackage{graphicx}
\usepackage{amssymb}
\usepackage{amsmath}
\usepackage{color}
\begin{document}

\title{ On the longitudinal structure function\\ in the dipole model}
\author{Marek Niedziela$^{(1),(2)}$, Michal\ Praszalowicz$^{(1)}$
\address{$^{(1)}$M. Smoluchowski Institute of Physics, Jagiellonian University, \\
S. {\L}ojasiewicza 11, 30-348 Krak{\'o}w, Poland. \\
$^{(2)}$Present address: Institute for Experimental Physics, University of Hamburg, \\
Luruper Chaussee 149,
D-22761 Hamburg, Germany.}}

%\thanks{\tt marek.niedziela@desy.de}
%\emailAdd{\tt michal@if.uj.edu.pl}

\maketitle

\begin{abstract}
We compare new HERA data for the longitudinal structure function $F_{\rm L}$
with the predictions of different variants of the dipole model. In particular we show that
the ratio $F_{\rm L}/F_2$ is  well described by the dipole models and is rather
insensitive to the details of the fit.
Fits to $F_2$ are performed  with the help of  geometrical scaling (GS).
Using the property of GS we 
derive the bounds for $F_{\rm L}/F_2$ both for the different versions of the dipole model
and in the general case. 
Finally we briefly discuss how the higher Fock components
of the photon wave function may affect these bounds.
\end{abstract}


\section{Introduction}

\begin{figure}[h!]
\centering
\includegraphics[width=6.0cm]{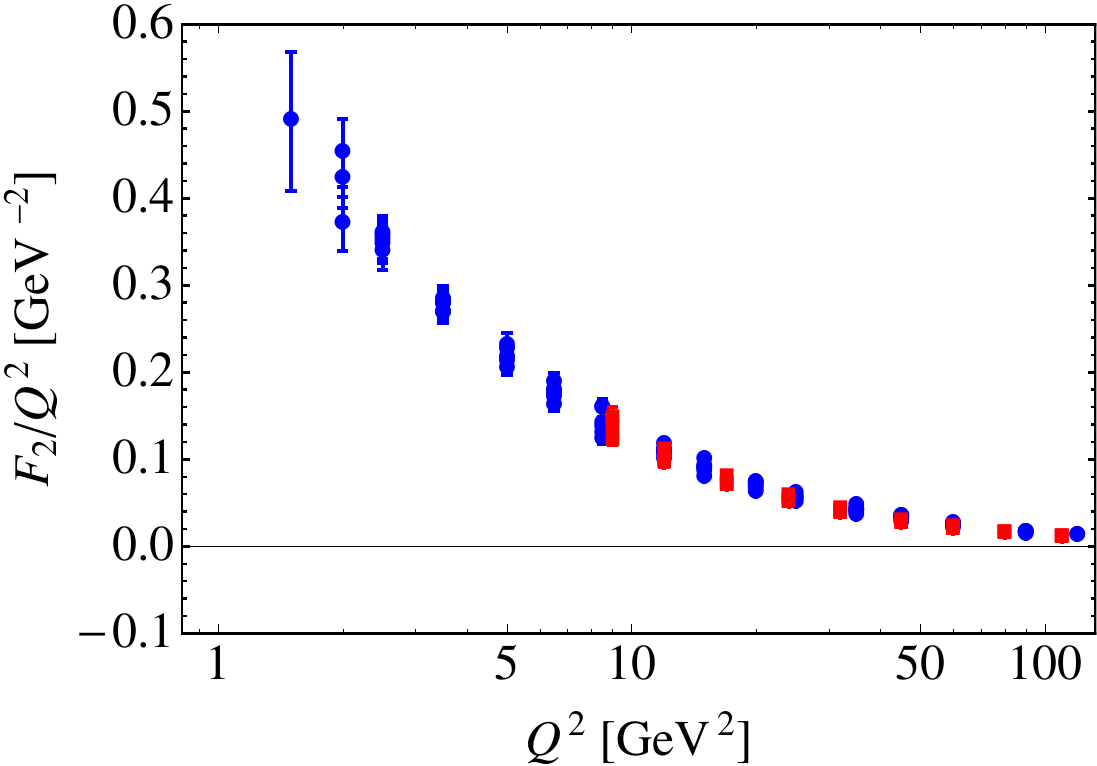}~
\includegraphics[width=6.0cm]{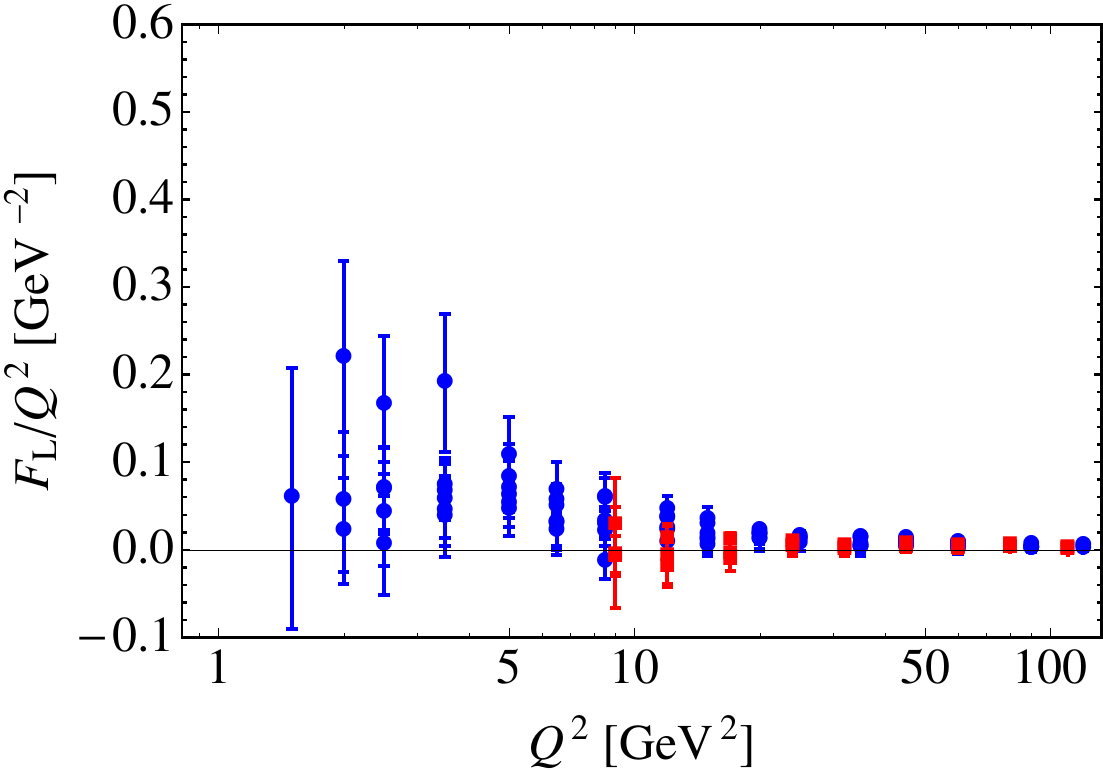}
\caption{H1
(blue circles) and ZEUS (red squares) data
\cite{Andreev:2013vha,Abramowicz:2014jak} for $F_{2}/Q^{2}$ (left) 
and $F_{\rm L}/Q^{2}$ (right) plotted as
functions of $Q^{2}$. Different points for one value of $Q^{2}$ correspond to
different Bjorken $x$'s.}%
\label{fig:F2L_data}%
\end{figure}

Recently H1 \cite{Andreev:2013vha} and ZEUS \cite{Abramowicz:2014jak}
Collaborations have published new data on the longitudinal structure function
$F_{\mathrm{L}}(x,Q^{2})$ in deep inelastic ep scattering (DIS). The H1
analysis extends and improves previous data \cite{Aaron:2008ad}, which now
cover kinematical range from $Q^{2}=1.5$~GeV$^{2}/c^{2}$ and $x=0.279\times
10^{-4}$ up to $Q^{2}=800$~GeV$^{2}/c^{2}$ and $x=0.0322.$ ZEUS\ data has been taken
in much smaller region from $Q^{2}=9$ up to $Q^{2}=110$~GeV$^{2}/c^{2} $ (see
\cite{Levy:2014tna} for summary). In both data sets there is strong
correlation between $Q^{2}$ and $x$ values; for each $Q^{2}$ structure function $F_{\mathrm{L}}$
(and also $F_2$ that has been measured in the same kinematical points)
has been measured over a limited $x$ range, with small $x$'s concentrated
around small values of $Q^{2}$, see  Fig.~\ref{fig:F2L_data}. 
Moreover, since $F_{\mathrm{L}}$ is difficult
to extract experimentally, even recent improved data has still large errors. 

Longitudinal structure function is of particular interest since, in the first
approximation of the parton model, it is equal identically zero
\cite{Callan:1969uq} (Callan-Gross relation) and therefore it is generated
entirely by radiative corrections. On the contrary, in the dipole model
$F_{\mathrm{L}}$ is nonzero, albeit small. Indeed, Nachtmann and collaborators
have shown that in the dipole model there exists a strict bound that
\cite{Ewerz:2006an, Ewerz:2007md,Ewerz:2012az}
\begin{equation}
F_{\mathrm{L}}\leq g_{\text{max}}\times F_{2}=0.27\times F_{2}\,.\label{bound}%
\end{equation}
This result, hereafter referred to as an EMNS bound, is independent of the
dipole-proton cross-section, and -- strictly speaking -- follows solely from the
properties of the photon-$\bar{q}q$ wave function.

Using previous H1 data \cite{Aaron:2008ad} the authors of Ref.~\cite{Ewerz:2012az} 
have shown that the bound (\ref{bound}) was
almost saturated, which is difficult to realize in realistic dipole models.
In this paper we are going to check if this conclusion remains still valid for
the new data and what is the value of bound (\ref{bound}) for commonly used
dipole-proton cross-sections. Similar analysis for the 
Golec-Biernat--W{\"u}sthoff model \cite{GolecBiernat:1998js,GolecBiernat:1999qd}
has been already performed in Ref.~\cite{Ewerz:2006vd}.
To simplify the analysis we use here the property of geometrical
scaling \cite{Stasto:2000er} (GS) which is to large precision exhibited by the
DIS data up to relatively large Bjorken $x$'s \cite{Praszalowicz:2012zh}. We
find that for realistic dipole-proton cross-sections the bound is indeed lower
than (\ref{bound}) with $g\simeq0.22$ ({\em i.e} 18.5\% below the bound 
of Eq.~(\ref{bound})).  In reality these bounds would be lowered if charm quark
mass was included.

With present experimental accuracy we do not find any significant 
tension between $F_{\rm L}/F_2$ data and the dipole model. Should such tension 
arise when new data appear, higher order corrections to the dipole model
might resolve the issue. Therefore  we discuss a possibility that corrections to the dipole
model coming from higher Fock states in the virtual photon may change
(\ref{bound}). Higher Fock components are needed \emph{e.g.} in the
dipole model \cite{GolecBiernat:1998js} description of the diffractive data
\cite{GolecBiernat:1999qd}. 
We show that this is possible only if there exists
a bound for higher Fock components that is analogous to (\ref{bound}), but
with longitudinal contribution to $F_{2}$ that is significantly different than in the case of the
$\bar{q}q$ state. Only explicit calculation of the $\bar{q}qg$ contribution to
the virtual photon wave function might give here a definite answer. Such calculations
have been recently carried out by various authors 
\cite{Balitsky:2010ze,Beuf:2011xd,Boussarie:2014lxa}, however these results have
not been so far applied to the phenomenological analysis of DIS. Although the calculation
of  $F^{\bar{q}qg}_{\rm L}/F^{\bar{q}qg}_2$ with the help of 
Refs.~\cite{Beuf:2011xd,Boussarie:2014lxa} might be probably possible, 
it is beyond the scope of the present paper.

The paper is organized as follows. In Sect.~\ref{sec:GSandbound} we rederive
the EMNS bound using geometrical scaling. Next, in Sect.~\ref{sec:dipmods},
we fit two versions of the dipole model to the present data set for $F_2$. To this end
we also use the property of geometrical scaling. We then
compare these fits with the data for $F_{\rm L}$ and discuss fit uncertainties.
In Sect.~\ref{sec:modelsbound} we calculate ratio $F_{\rm L}/F_2$ for the 
afore mentioned fits and compare it with the data and with the EMNS bound. We do not find
large tension between the data and model predictions. An influence of higher
Fock states on the EMNS bound is discussed in Sect.~\ref{sec:HFS}. We conclude
in Sect.~\ref{sec:conclusions}.

\section{Geometrical scaling and the EMNS bound}
\label{sec:GSandbound}

For three massless flavors DIS structure functions read  \cite{empty}:%
\begin{align*}
F_{2}(x,Q^{2})  & =\frac{Q^{2}}{4\pi^{3}}\int dr^{2}\left\{  \left\vert
\psi_{\rm T}(r,Q^{2})\right\vert ^{2}+\left\vert \psi_{\rm L}(r,Q^{2})\right\vert
^{2}\right\}  \,\sigma_{\text{dp}}(r^{2}),\\
F_{\mathrm{L}}(x,Q^{2})  & =\frac{Q^{2}}{4\pi^{3}}\int dr^{2}\left\vert
\psi_{\rm L}(r,Q^{2})\right\vert ^{2}\,\sigma_{\text{dp}}(r^{2})
\end{align*}
where photon wave functions take the following form%
\begin{align}
|\psi_{\rm T}(r,Q^{2})|^{2}  & =\int_{0}^{1}dz\left[  z^{2}+(1-z)^{2}\right]
\overline{Q}^{2}K_{1}^{2}(\overline{Q}r),\nonumber\\
|\psi_{\rm L}(r,Q^{2})|^{2}  & =4\int_{0}^{1}dz\,z(1-z)\overline{Q}^{2}K_{0}%
^{2}(\overline{Q}r)\, .
\end{align}
Here $K_{i}$ are modified Bessel functions and
\begin{equation}
\overline{Q}^{2}=z(1-z)Q^{2}.
\end{equation}
It is convenient to define functions $\Phi_{\rm T,L}$%
\begin{equation}
\Phi_{\rm T,L}(u=rQ)=r^{2}\left\vert \psi_{\rm T,L}(r,Q^{2})\right\vert ^{2}%
\end{equation}
that depend only on the combined variable $u=Qr$. Therefore%
\begin{align}
F_{2}(x,Q^{2})  & =\frac{Q^{2}}{2\pi^{3}}\int du\left\{  \Phi_{\rm T}(u)+
\Phi_{\rm L}(u)\right\}  \frac{\sigma_{\text{dp}}(u/Q)}{u},\nonumber\\
F_{\mathrm{L}}(x,Q^{2})  & =\frac{Q^{2}}{2\pi^{3}}\int du\,\Phi_{\rm L}%
(u)\frac{\sigma_{\text{dp}}(u/Q)}{u}.\label{F2FL}%
\end{align}
This parametrization is very convenient for the following reasons. First of
all wave functions $\Phi_{\rm T,L}(u)$ are universal and do not depend on external
kinematical variables. Secondly, unlike functions $\psi_{\rm T,L}(r,Q^{2})$, they
are everywhere regular in $u$. And finally, cross-section $\sigma_{\rm dP}(u/Q)/u$
is a localized function of variable $u$ that vanishes both for $u\rightarrow0$
and $u\rightarrow\infty$.

If -- as it is in the case of the GBW model -- the dipole-proton cross-section
exhibits geometrical scaling, \emph{i.e.} $\sigma_{\rm dP}(r)=\sigma
_{\rm dP}(rQ_{\text{s}}(x))$ then the integral%
\begin{equation}%
%TCIMACRO{\dint }%
%BeginExpansion
{\displaystyle\int}
%EndExpansion
du\,\Phi_{\rm T,L}(u)\frac{\sigma_{\text{dp}}\left(  u/Q\right)  }{u}=%
%TCIMACRO{\dint }%
%BeginExpansion
{\displaystyle\int}
%EndExpansion
du\,\Phi_{\rm T,L}(u)\frac{\sigma_{\text{dp}}\left(  u\,Q_{\text{s}}/Q\right)
}{u}=\text{function}(\tau)
\end{equation}
depends only on a scaling variable%
\begin{equation}
\tau=\left(  Q/Q_{\text{s}}\right)  ^{2}.
\end{equation}
Here $Q_{\text{s}}^{2}$ is a saturation scale%
\begin{equation}
Q_{\text{s}}^{2}=Q_{0}^{2}\left(  \frac{x}{x_{0}}\right)  ^{-\lambda
}.\label{Qsat}%
\end{equation}

\begin{figure}[t]
\centering
\includegraphics[width=6.5cm]{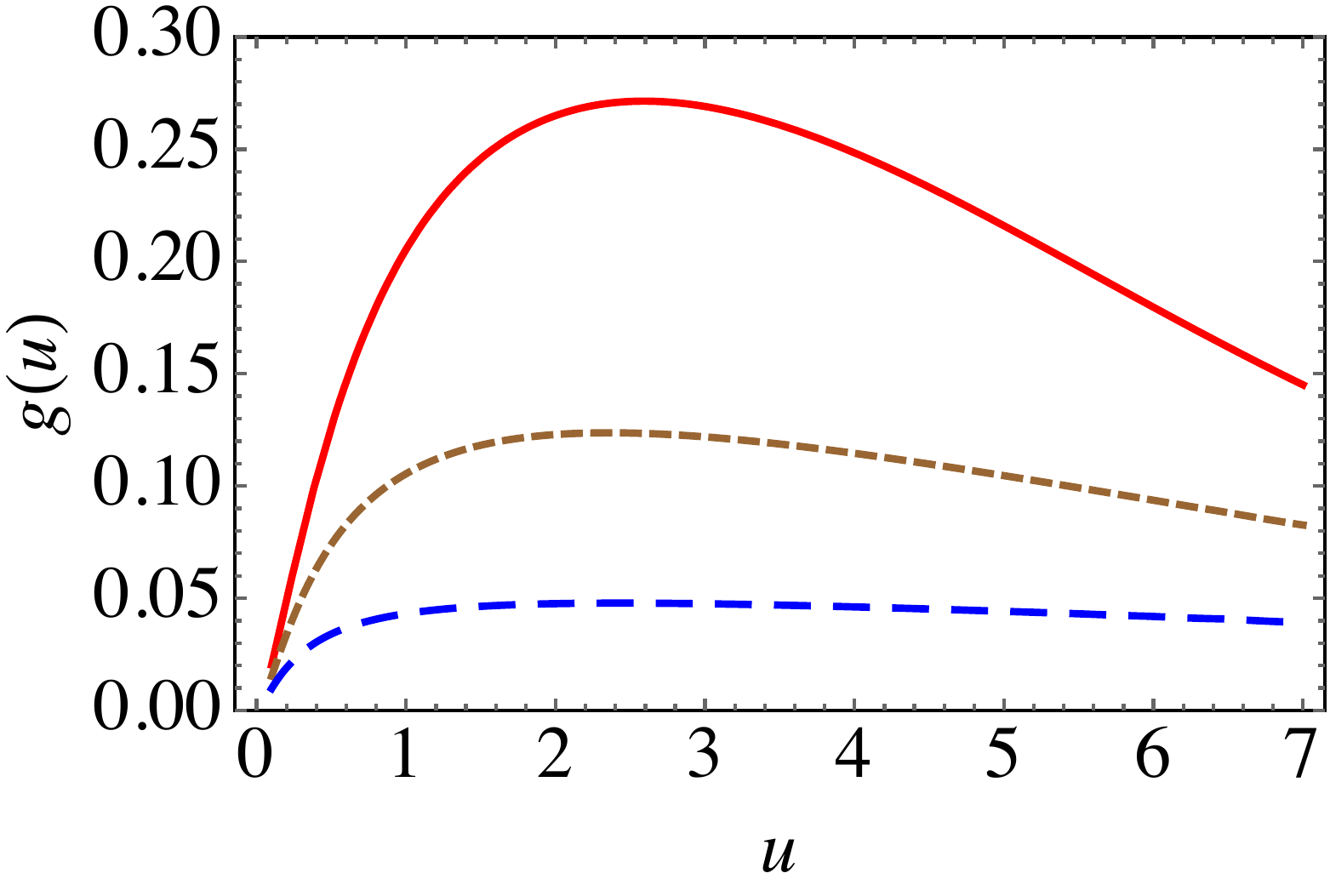}\caption{Solid (red) line: function $g(u)$ defined in
Eq.~(\ref{defgu}). Short dashed (brown) line: function $g(u)$ in the case of  massive quarks for
$\delta_{\rm f}=m_{\rm f}/Q=0.5$ and long dased (blue) line:  for
$\delta_{\rm f}=m_{\rm f}/Q=1.06$, which corresponds to the charm quark mass $m_{\rm c}=1.3$~GeV/$c$
and  $Q^2=1.5$~GeV$^2/c^2$.
}%
\label{fig:gu}%
\end{figure}

Now we can rederive the EMNS bound by considering the ratio%
\begin{equation}
\frac{F_{\mathrm{L}}(\tau)}{F_{2}(\tau)}=\frac{%
%TCIMACRO{\dint }%
%BeginExpansion
{\displaystyle\int}
%EndExpansion
du\,\Phi_{\rm L}(u)\,\sigma_{\text{dp}}(u/Q)/u}{%
%TCIMACRO{\dint }%
%BeginExpansion
{\displaystyle\int}
%EndExpansion
du\left\{  \Phi_{\rm T}(u)+\Phi_{\rm L}(u)\right\}  \,\sigma_{\text{dp}}(u/Q)/u}%
\end{equation}
and observing that function%
\begin{equation}
g(u)=\frac{\Phi_{\rm L}(u)}{\Phi_{\rm T}(u)+\Phi_{\rm L}(u)},\label{defgu}%
\end{equation}
which is plotted in Fig.~\ref{fig:gu}, has a maximum $g_{\text{max}}=0.2714$
for $u=2.591$. Therefore it follows that
\begin{equation}
\frac{F_{\mathrm{L}}(x,Q^{2})}{F_{2}(x,Q^{2})}\leq g_{\mathrm{max}%
}=0.27.\label{Nbound}%
\end{equation}

The bound (\ref{Nbound}) has been derived for the case of massless quarks.
While this is certainly a good approximation for three light flavors, given the
fact that the lowest photon virtuality in the data set we use is
$Q^{2}=1.5$~GeV$^{2}/c^{2}$, the inclusion of charm mass effects is going
to change (\ref{Nbound}). For a given flavor function $g$ defined in Eq.~(\ref{defgu})
is no longer a function of scaling variable $u$ only
but in addition depends on the ratio $\delta_{\rm f}^2=m_{\rm f}^2/Q^2$.
For large $Q^2$ ({\em i.e.} small  $\delta_{\rm f}$) $g(u,\delta_{\rm f}) \rightarrow g(u)$.
Moreover we have found numerically that everywhere in $u$ we have
\begin{equation}
g(u,\delta_{\rm f}) \le g(u)
\end{equation}
and the maxima $g^{({\rm f})}_{\rm max}$ of $g(u,\delta_{\rm f})$ are decreasing with increasing $\delta_{\rm f}$,
as illustrated in Fig.~\ref{fig:gu}. This is consistent with the observation of
Ref.~\cite{Ewerz:2006an} that $g^{({\rm f})}_{\rm max}$ is a monotonically growing function of $Q^2$.
Therfore
\begin{equation}
0 \le g^{({\rm f})}_{\rm max}(Q^2) \le g_{\rm max}\, .
\label{gfbound}
\end{equation}
This  allows us to estimate the effect of the charm quark on the ratio:
\begin{eqnarray}
\frac{F^{({\rm light+c})}_{\rm L}}{F^{({\rm light+c})}_2}& = & 
\frac{F_{\rm L}+F^{({\rm c})}_{\rm L}}{F_{2}+F^{({\rm c})}_{2}} =
\frac{{F_{\rm L}}/{F_2}+F^{({\rm c})}_{\rm L}/F^{({\rm c})}_{2}\, F^{({\rm c})}_{2}/{F_2}}{1+F^{({\rm c})}_{2}/{F_2}}
\nonumber \\
& \le & g_{\rm max}\frac{1 +g^{({\rm c})}_{\rm max}/g_{\rm max}\, F^{({\rm c})}_{2}/{F_2}}{1+F^{({\rm c})}_{2}/{F_2}}
\le g_{\rm max} 
\end{eqnarray}
where the last inequality follows from (\ref{gfbound}). Note that $F_{2,\,{\rm L}}$ without any superscript refers to the
structure function with light flavors only and that dependence on $Q^2$ has been suppressed.
We see therefore, in agreement with Ref.~\cite{Ewerz:2012az}, that
inclusion of charm lowers the bound on $F^{({\rm light+c})}_{\rm L}/F^{({\rm light+c})}_2$ 
in proportion that depends on $F_2^{({\rm c})}/F_2$. For $m_{\rm c}=1.3$~GeV/$c$ and for the lowest
$Q^2$ in the present data set we get numerically $g^{({\rm c})}_{\rm max} \approx 0.05$ (see Fig.~(\ref{fig:gu})), which
gives $g^{({\rm c})}_{\rm max}/g_{\rm max} \approx 0.19$. We do not know what is the fraction
of charm  in the present data sample, however taking a typical value of 
$F^{({\rm c})}_{2}/{F_2} \sim 25\%$, we get that $g_{\rm max}^{({\rm light + c})} \sim 0.23$.
This means that bound (\ref{Nbound}) is lowered for the lowest $Q^2$ bin by approximately 18\%
and is approaching (\ref{Nbound}) for higher $Q^2$.
In the following we will ignore charm contribution and stick to the bound (\ref{Nbound}) coming
back to this point in Sect.~\ref{sec:modelsbound}.

\section{Dipole models and geometrical scaling}
\label{sec:dipmods}

In order to check how far the bound (\ref{Nbound}) overestimates actual
predictions of the dipole model with realistic dipole-proton cross-section,
we are going  to compute ratio (\ref{Nbound}) for a given $\sigma_{\text{dp}}$ in
terms of scaling variable $\tau$. We will see that for commonly used
parametrizations of $\sigma_{\text{dp}}$, ratio $F_{\mathrm{L}}/F_{2}$ is a
slowly varying function of $\tau$ with a maximum equal approximately $0.216 - 0.224$,
which only slightly depends on the parametrization actually used. To this end
we have decided to perform our own fits to the $F_{2}$ data over the
restricted kinematical range where the longitudinal structure function
$F_{\mathrm{L}}$ has been measured by H1. The reason for this is threefold.
Firstly, new data is of much better quality than the previous data to which
the dipole model parameters have been fitted. Secondly, we do not aim at a
global fit, but rather at a fit which covers only the points where
$F_{\mathrm{L}}$ has been measured as well. Therefore fit parameters -- as
we shall see in the following -- will be different from the ones obtained in
the global fits. And finally, we have decided to fit the data with the help of
geometrical scaling -- a procedure not used so far in the fits to the DIS data.

\begin{table}[t]
\centering
%%%table corrected%%%%%
\begin{tabular}
[c]{lcccc}%
$x_{\text{max}}$ & $\sigma_{0}[1/$GeV$^{2}$] & $\lambda$ & $x_{0}$ & $\chi
^{2}/$dof\\\hline
none & 23.68 & 0.389 & 0.010497 & 1.18\\
0.01 & 27.11 & 0.353 & 0.007786 & 0.87\\
0.005 & 29.33 & 0.333 & 0.006435 & 0.79\\
0.0005 & 38.37 & 0.253 & 0.003090 & 0.70\\\hline
\end{tabular}
\caption{Parameters of the GBW model fitted to $F_{2}$ H1 data up to
$x_{\mathrm{max}}$.}%
\label{tab:GBW}%
\end{table}

Fitting dipole models to the data becomes very easy when $F_{2}$ depends only
upon single scaling variable $\tau$. This happens because points corresponding
to one particular value of $Q^{2}$ but different $x$'s (see "stacks" in
Fig.~\ref{fig:F2L_data}) correspond to different values of $\tau$ and are
therefore shifted horizontally -- if plotted in terms of $\tau$ -- by values
that are different for different $x$'s. As a consequence dipole model predictions
fall on a universal curve (up to an overall normalization $\sigma_{0}$), and
data fitting consists in changing $\sigma_{0}$ and the parameters defining
scaling variable $\tau$, \emph{i.e.} $x_{0}$ and $\lambda$. By varying these
three parameters one forces experimental points to fall on theoretical
prediction, rather than by changing theoretical predictions one is trying to
reproduce experimental points. Therefore this method is very efficient, as it
does not require time consuming recalculations of the theoretical curve.

Even in the case of dipole-proton cross-sections that violate  GS by
explicit (albeit weak) dependence on $x$, like in the model
of Iancu, Itakura and Munier \cite{Iancu:2003ge}, it is still possible to make a GS-like fit by
keeping $x$ at some fixed average value $x_{\mathrm{ave}}$ and then study the
uncertainty  of theoretical predictions by changing $x$ over the range that is
covered by experimental data. We shall come back to this point later.

\begin{figure}[h!]
\centering
\includegraphics[width=6.0cm]{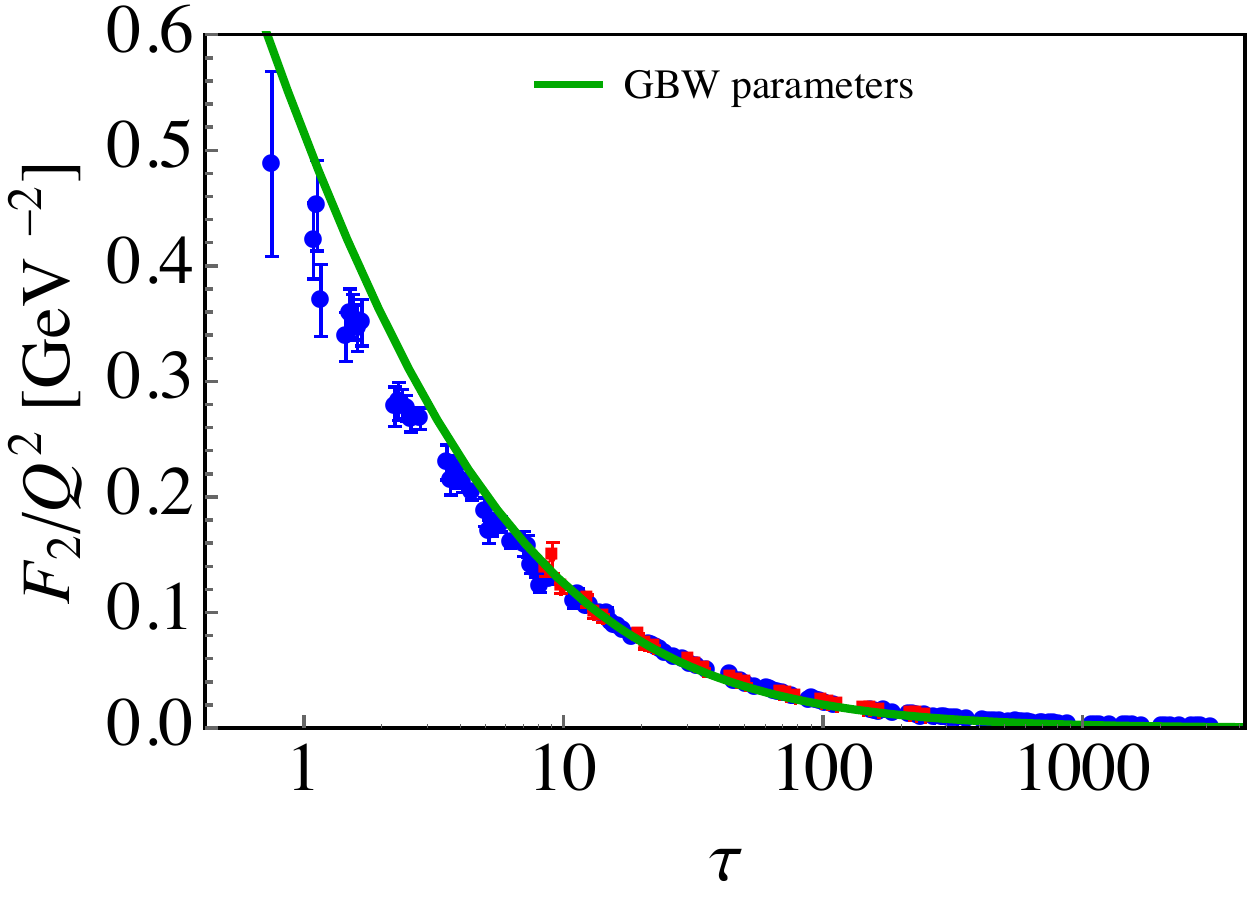}~~
\includegraphics[width=6.0cm]{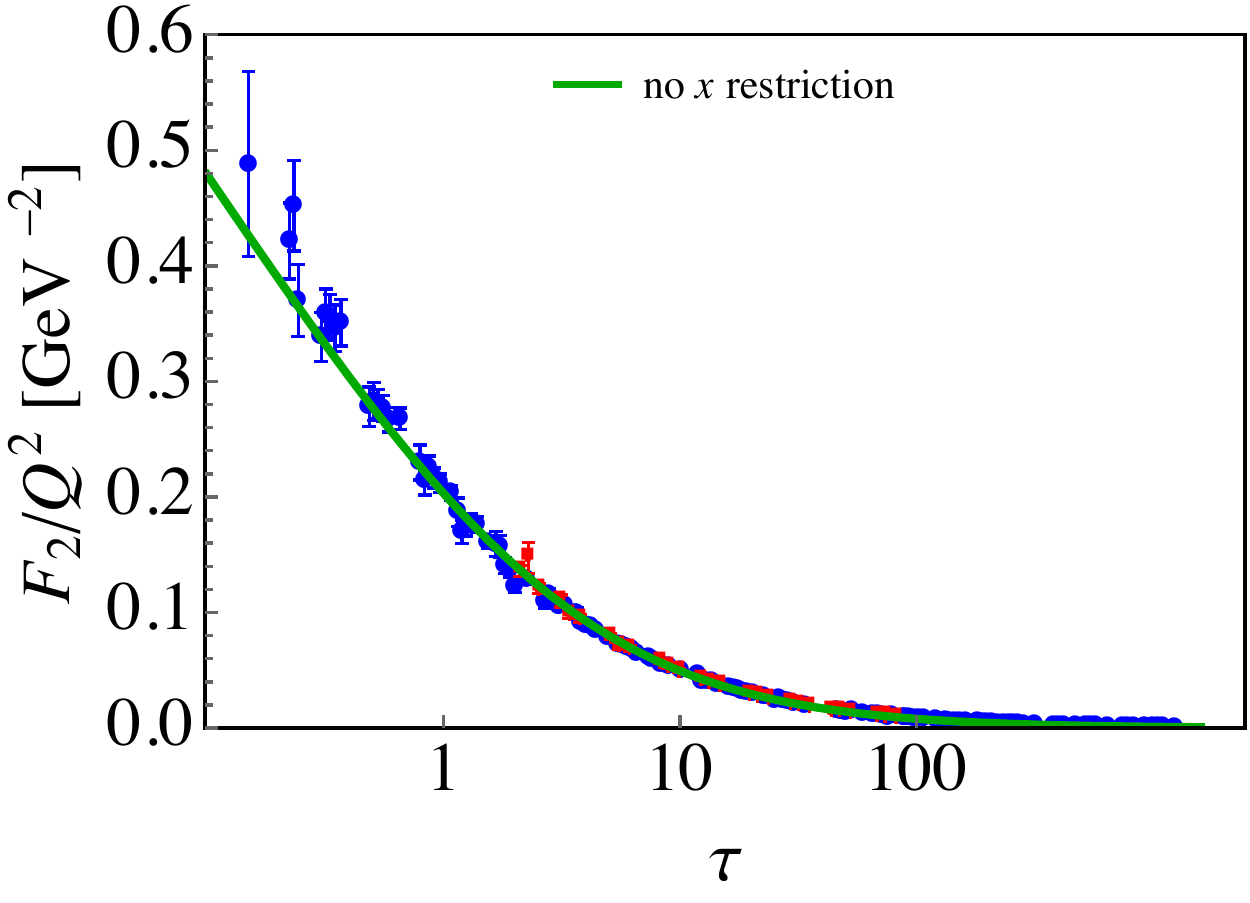}\\%
\includegraphics[width=6.0cm]{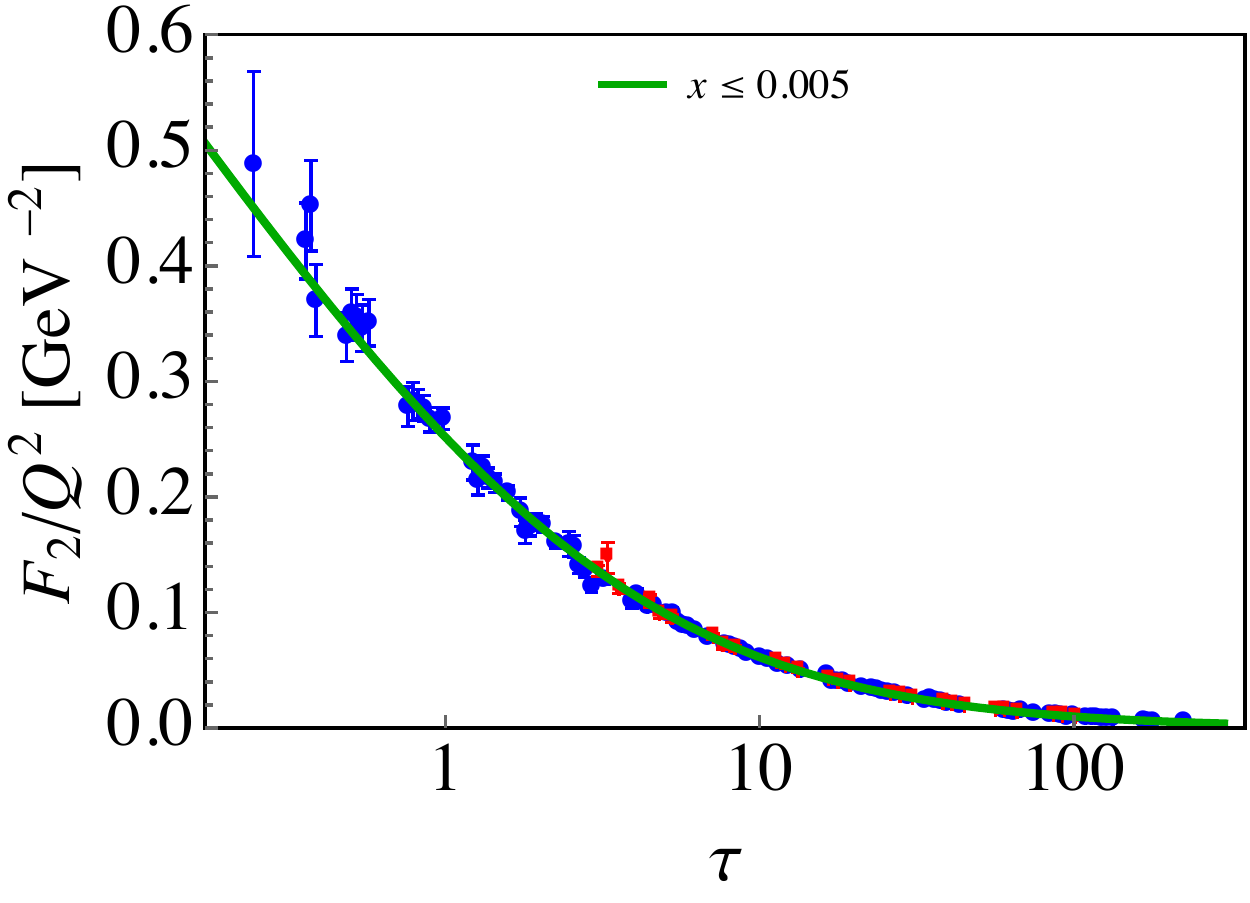}~~
\includegraphics[width=6.0cm]{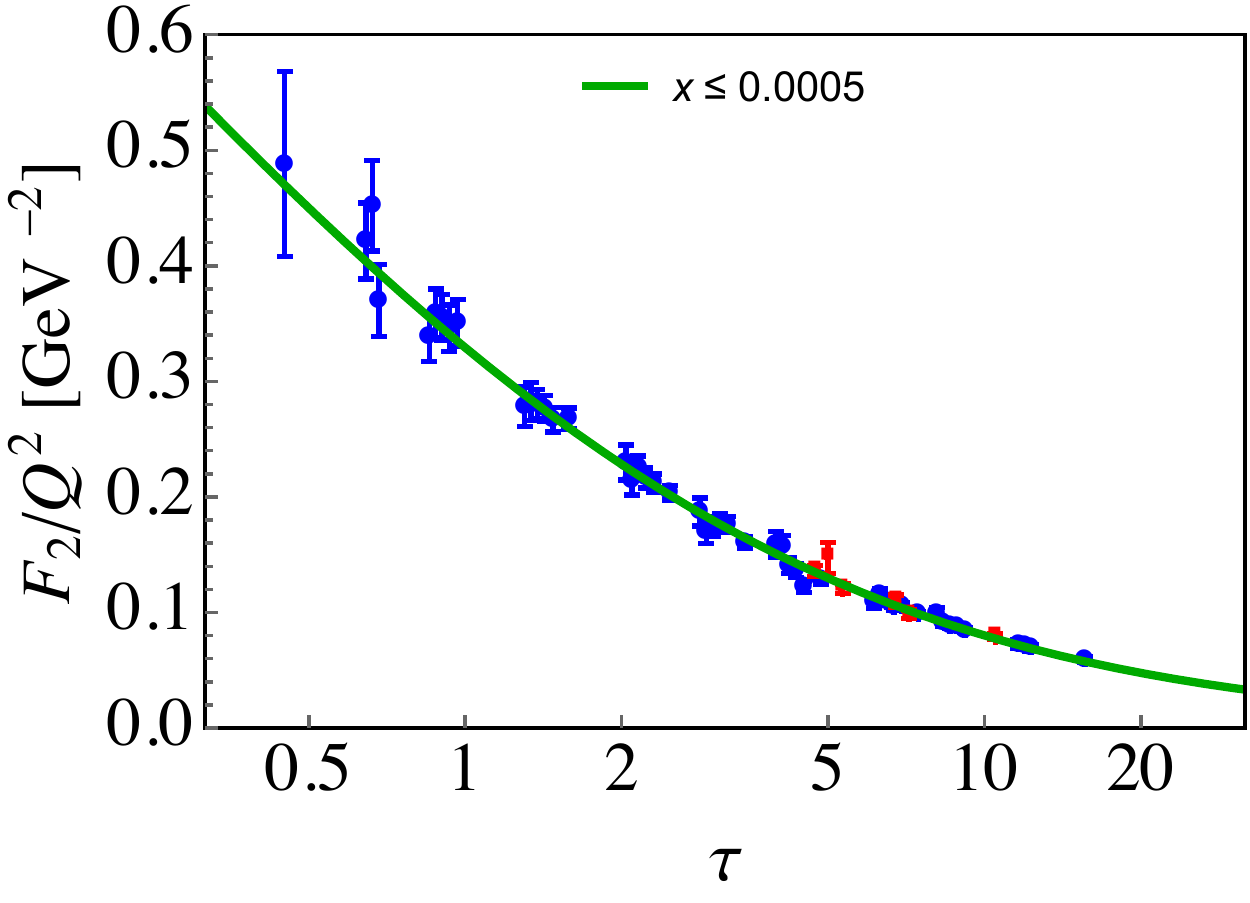}
\caption{H1
(blue circles) and ZEUS (red squares) data
\cite{Andreev:2013vha,Abramowicz:2014jak}
for $F_{2}/Q^{2}$ plotted as a function of scaling variable $\tau$ for
different values of fit parameters corresponding to the GBW model 
given in Table~\ref{tab:GBW}. Fit
parameters in the first panel correspond to the original fit of
Ref.~\cite{GolecBiernat:1998js} with no charm data included.}%
\label{fig:F2_GBW}%
\end{figure}

Let us first consider the simplest version of the dipole model, namely the GBW
parametrization \cite{GolecBiernat:1998js}, 
which -- up to an overall normalization $\sigma_{0}$ -- takes
the following form in terms of scaling variable $\tau$%
\begin{equation}
\frac{\sigma_{\text{dp}}^{\text{GBW}}(u,\tau)}{\sigma_{0}}=1-\exp\left(
-\frac{u^{2}}{4\tau}\right)  .\label{sigmaGBW}%
\end{equation}
Pluging (\ref{sigmaGBW}) into Eqs.(\ref{F2FL}) gives unnormalized theoretical
prediction for the structure functions divided by $Q^2$, which will be denoted in the following
by small characters $f_{2,L}(\tau)$. Experimental data are tabulated in a set
of discrete points $\{Q_{i}^{2},x_{i}\}$, and we fit three free parameters of
the model, $\sigma_{0},x_{0}$ and $\lambda$, by transforming experimental
entries in the following way:%
\begin{equation}
F_{2}(x_{i},Q_{i}^{2})\rightarrow\frac{1}{Q^2 \sigma_{0}}F_{2}(x_{i},Q_{i}%
^{2})=f_{2}\left(  \tau_{i}=\frac{Q^{2}}{Q_{\text{s}}^{2}(x_{i})}\right)
\end{equation}
and demanding that they are equal to the theoretical prediction at the
pertinent value of scaling variable $\tau_{i}$ with $Q_{\text{s}}$ given by
Eq.(\ref{Qsat}). The results are shown in Table~\ref{tab:GBW}. Since GS is
supposed to work the best for small values of Bjorken $x$'s we have performed
a number of fits restricting the allowed $x$ region up to a maximal value
denoted as $x_{\text{max}}$. We see that even without any cut on the maximal
value of $x$, \emph{i.e.} for $x$ as large as 0.0322 (the highest $x$ in the
analyzed data set) $\chi^{2}$ of the fit is quite reasonable. By restricting
analyzed data to the smaller and smaller range of Bjorken $x$'s $\chi^{2}$ is
getting smaller, but also model parameters vary substantially. Parameter
$\sigma_{0}$ is much smaller than in the original fit of
Ref.~\cite{GolecBiernat:1998js} $\sigma_{0}^{\mathrm{GBW}}= 23\;\mathrm{mb}%
=59\;\mathrm{GeV}^{-2}$. Exponent $\lambda$ approaches the value of
Ref.~\cite{GolecBiernat:1998js} $\lambda^{\mathrm{GBW}}=0.288$ only for small
$x_{\mathrm{max}}$ (note that maximal $x$ in Ref.~\cite{GolecBiernat:1998js}
was equal to 0.01, whereas the lowest $x=6\times10^{-6}$ was much below the
minimal $x$ of present analysis), otherwise being consistent with
model-independent analysis of Ref.~\cite{Praszalowicz:2012zh}. The results of
the fits, together with the original parametrization of
Ref.~\cite{GolecBiernat:1998js} are plotted in Fig.~\ref{fig:F2_GBW}. One can
see rather good agreement of fits from Table~\ref{tab:GBW} with the data, and
-- also quite importantly -- good quality of GS of the data.

\begin{figure}[h]
\centering
\includegraphics[width=6.0cm]{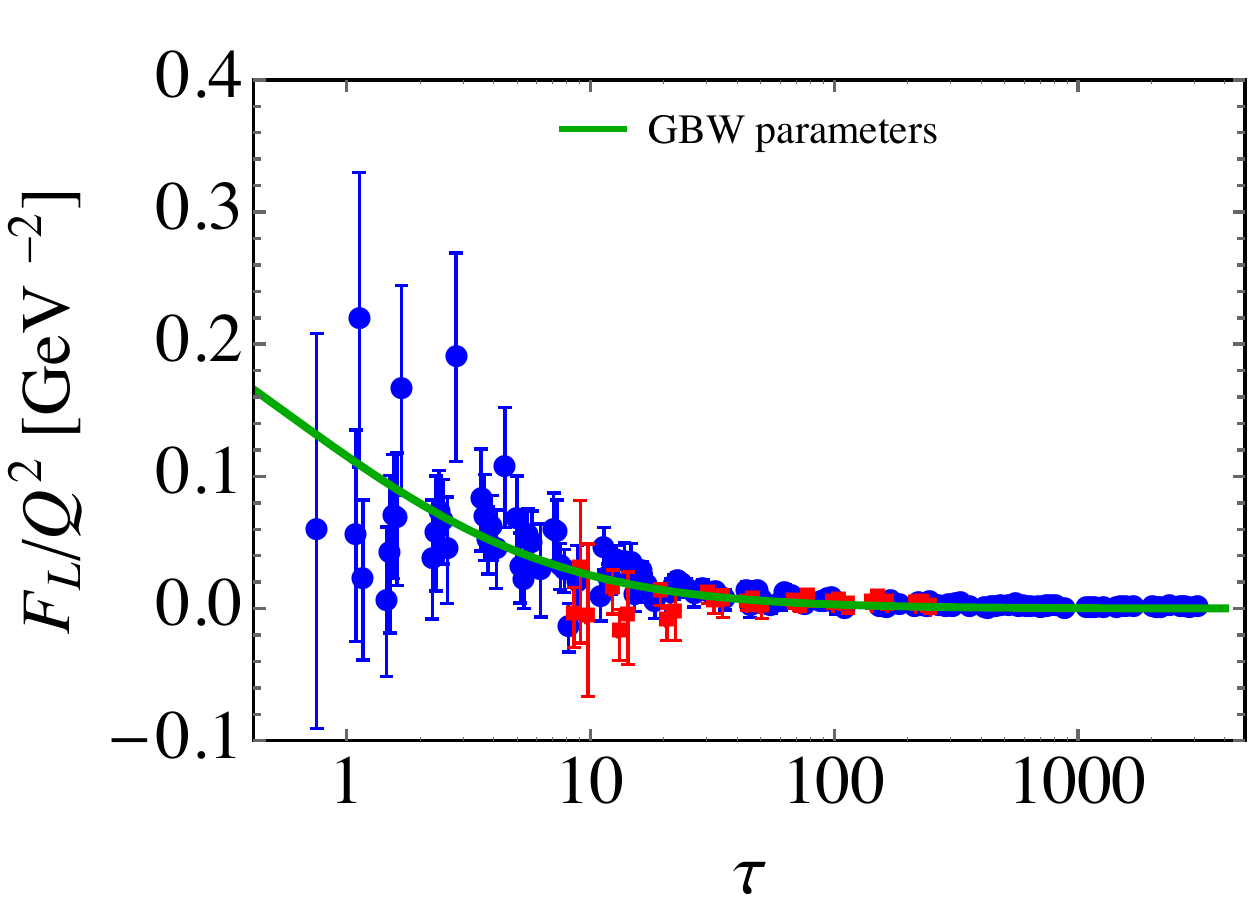}~
\includegraphics[width=6.0cm]{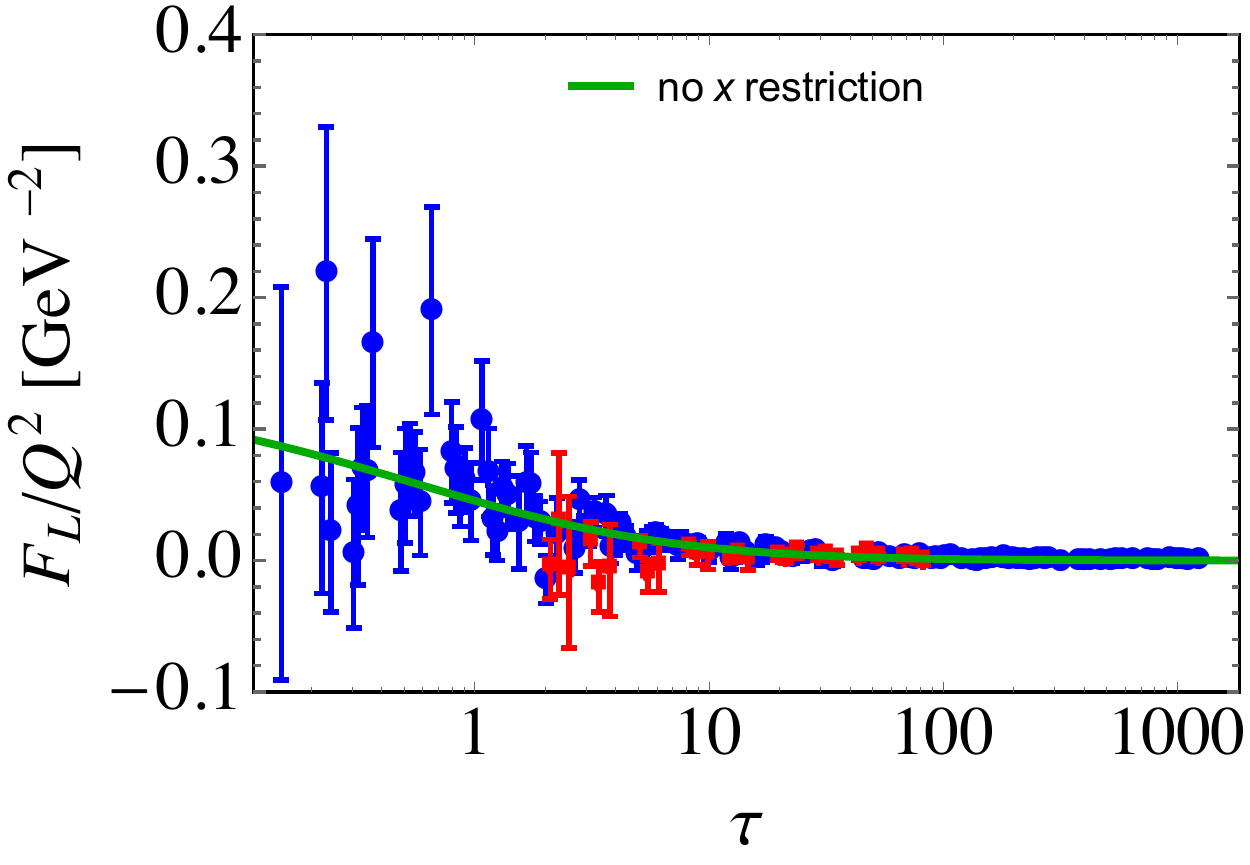}\\%
\includegraphics[width=6.0cm]{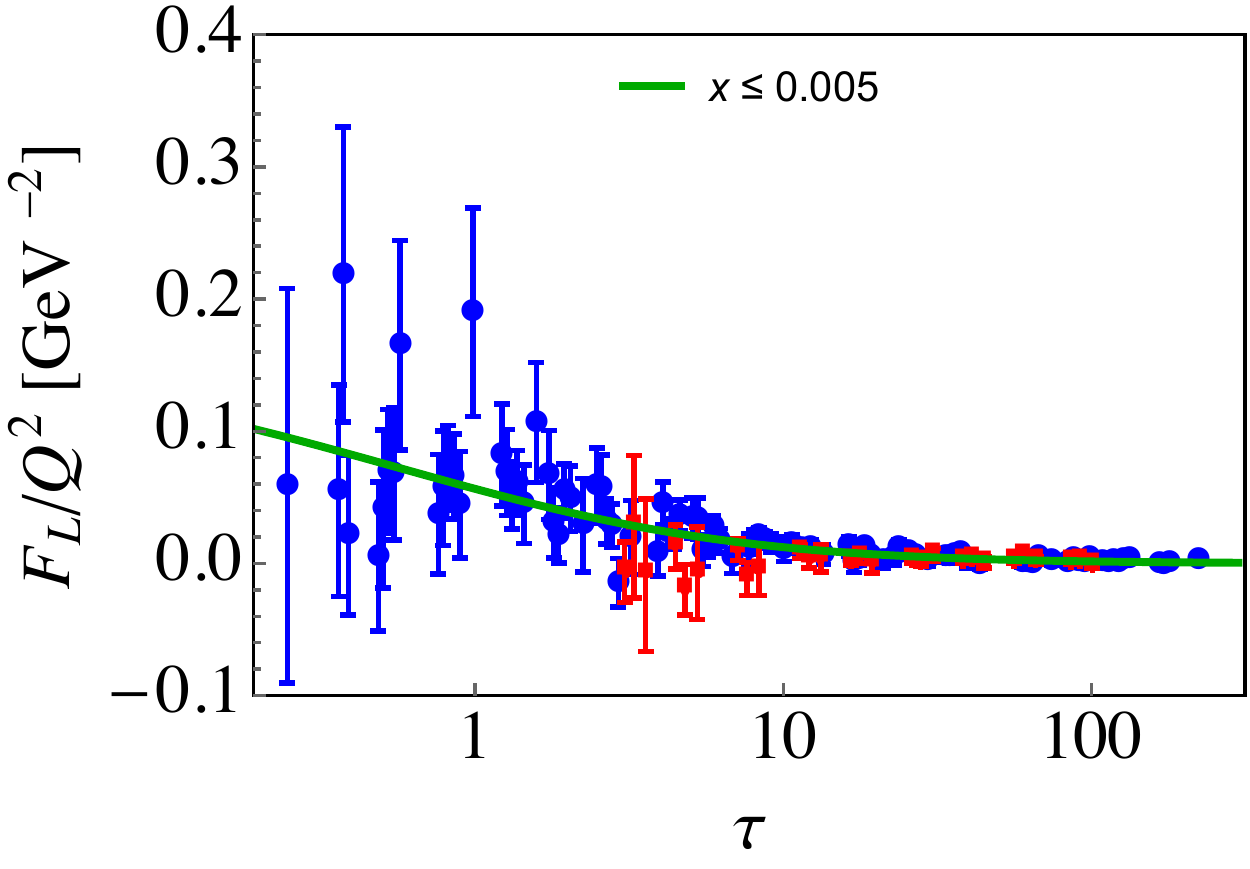}~
\includegraphics[width=6.0cm]{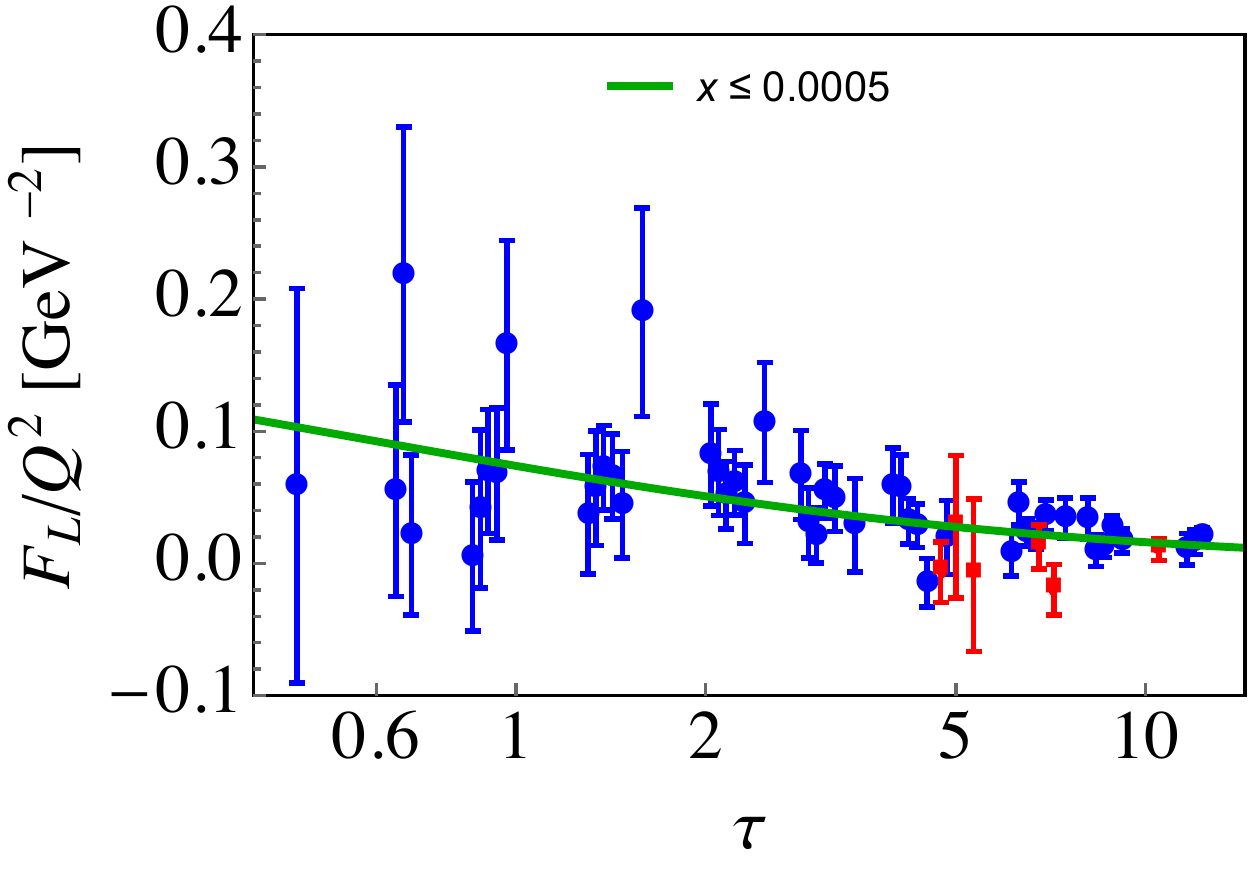}
\caption{H1
(blue circles) and ZEUS (red squares) data
\cite{Andreev:2013vha,Abramowicz:2014jak}  for $F_{\mathrm{L}}/Q^{2}$
plotted as a function of scaling variable $\tau$. Solid curves correspond to
the GBW model fits of Table~\ref{tab:GBW}. }%
\label{fig:FL_GBW}%
\end{figure}

Finally in Fig.~\ref{fig:FL_GBW} we plot data for $F_{\mathrm{L}}/Q^{2}$ as a
function of $\tau$ together with theoretical parametrization of
Ref.~\cite{GolecBiernat:1998js} and the present fits corresponding to 
 Tab.~\ref{tab:GBW}. We can see that due to still large experimental
errors of $F_{\mathrm{L}}$ all parametrizations, although different, describe
well the data.

As a second example let us consider a dipole model by Iancu, Itakura and
Munier (IIM) \cite{Iancu:2003ge}, where the dipole-proton cross-section is
defined in terms of two functions%
\begin{align}
A_{1}(u,\tau,x)  & =A_{0}\left(  \frac{u^{2}}{4\tau}\right)  ^{\gamma
+\frac{\ln(4\tau/u^{2})}{2\kappa\lambda\ln(1/x)}},\nonumber\\
A_{2}(u,\tau)  & =1-\exp\left(  -a\ln^{2}(bu/\sqrt{\tau})\right)
\label{A1A2_IIM}%
\end{align}
and
\begin{equation}
\frac{\sigma_{\text{dp}}^{\text{IIM}}(u,\tau,x)}{\sigma_{0}}=\left\{
\begin{array}
[c]{rrr}%
A_{1}(u,\tau,x) & \text{for} & u^{2}<4\tau\\
&  & \\
A_{2}(u,\tau) & \text{for} & 4\tau\leq u^{2}%
\end{array}
\right.  .\label{sigmaIIM}%
\end{equation}
Here $\gamma=0.63$ is related to the anomalous dimension of the forward
scattering amplitude in the BFKL formalism, while $\kappa=9.9$ corresponds to
the diffusion coefficient. Parameters $a$ and $b$ are determined uniquely by
gluing $A_{1}$ and $A_{2}$ and their derivatives at $u^{2}=4 \tau$. Parameter
$A_{0}$ is in principle free, however, as it was shown in
Ref.~\cite{Iancu:2003ge} the best $\chi^{2}$ was obtained for $A_{0}=0.7$ and
for the purpose of the present work we will keep it fixed at this value.
Therefore the only free parameters are, as in the case of the GBW model, an
overall normalization $\sigma_{0}$ and two parameters of the saturation scale:
$\lambda$ and $x_{0}$.

\begin{table}[t]
\centering
%%%table corrected%%%%%
\begin{tabular}
[c]{llcccc}%
$x_{\text{max}}$ & $x_{\text{ave}}$ & $\sigma_{0}[1/$GeV$^{2}]$ & $\lambda$ &
$x_{0}$ & $\chi^{2}/$dof\\\hline
none & 0.00359 & 20.22 & 0.597 & 0.002553 & 1.76\\
0.01 & 0.00182 & 21.50 & 0.583 & 0.002140 & 1.57\\
0.005 & 0.00121 & 25.56 & 0.531 & 0.001392 & 1.31\\
0.0005 & 0.00022 & 34.30 & 0.389 & 0.000645 & 0.75\\\hline
\end{tabular}
\caption{Parameters of the IIM \cite{Iancu:2003ge} model fitted to $F_{2}$ H1
data up to $x_{\mathrm{max}}$.}%
\label{tab:IIM}%
\end{table}

\begin{figure}[h!]
\centering
\includegraphics[width=6.0cm]{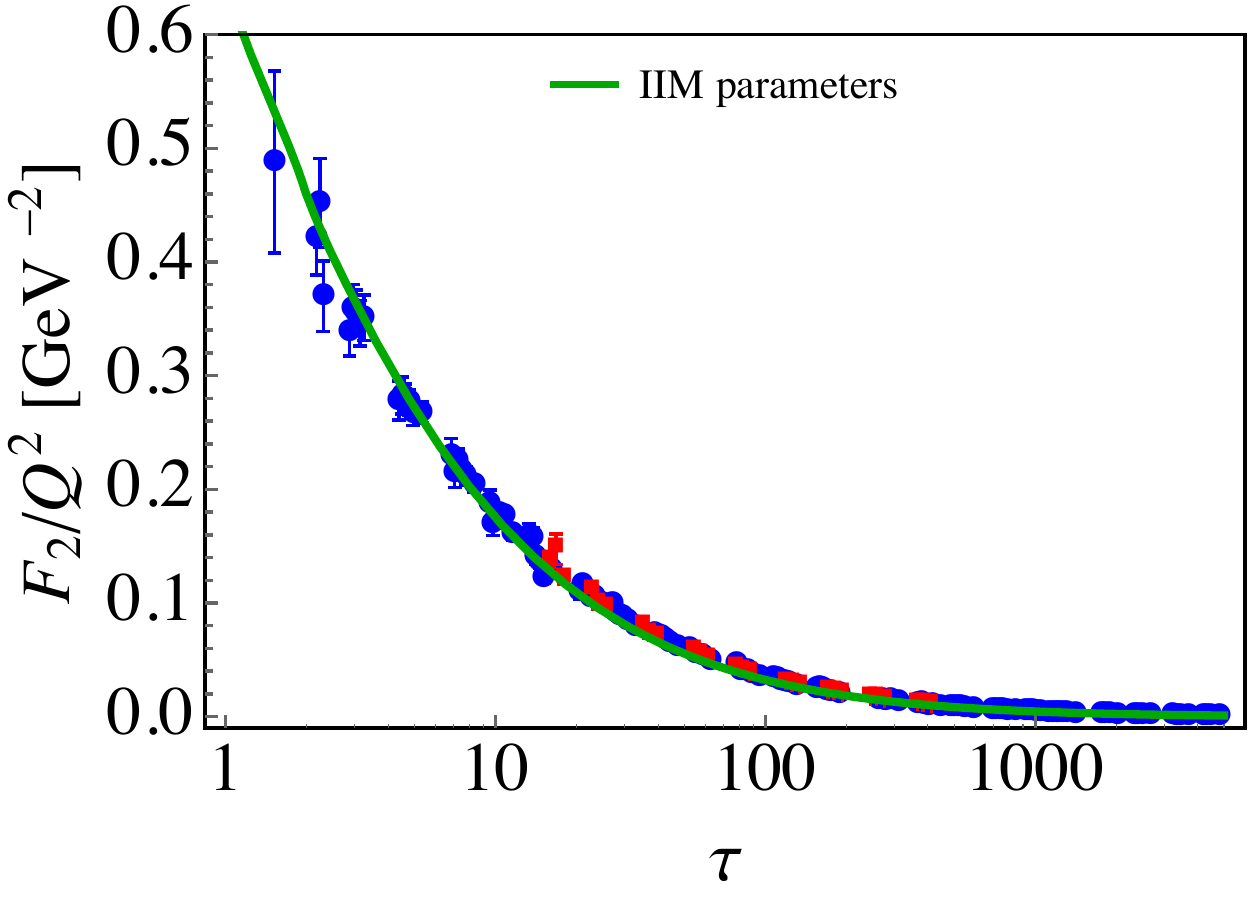}~\includegraphics[width=6.0cm]{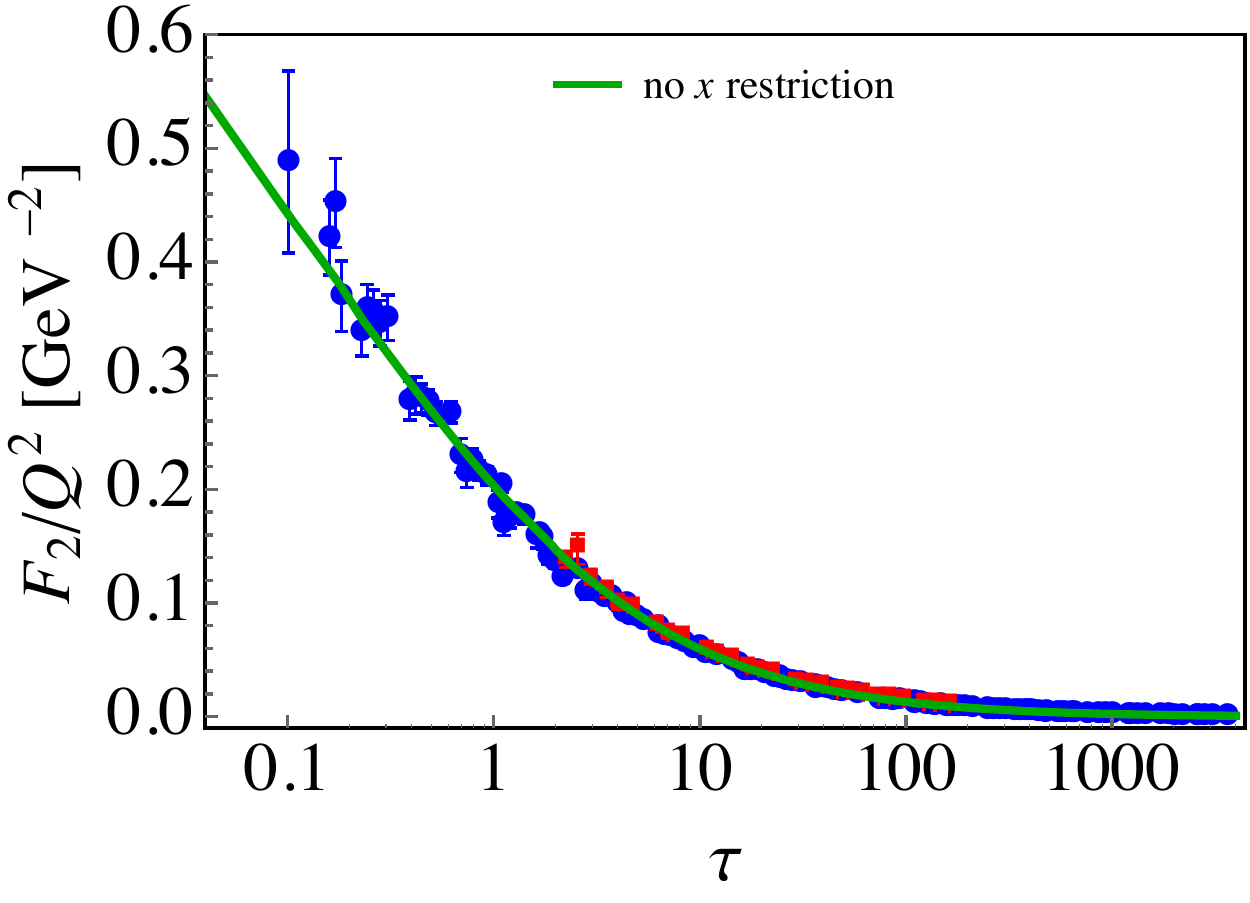}\\%
\includegraphics[width=6.0cm]{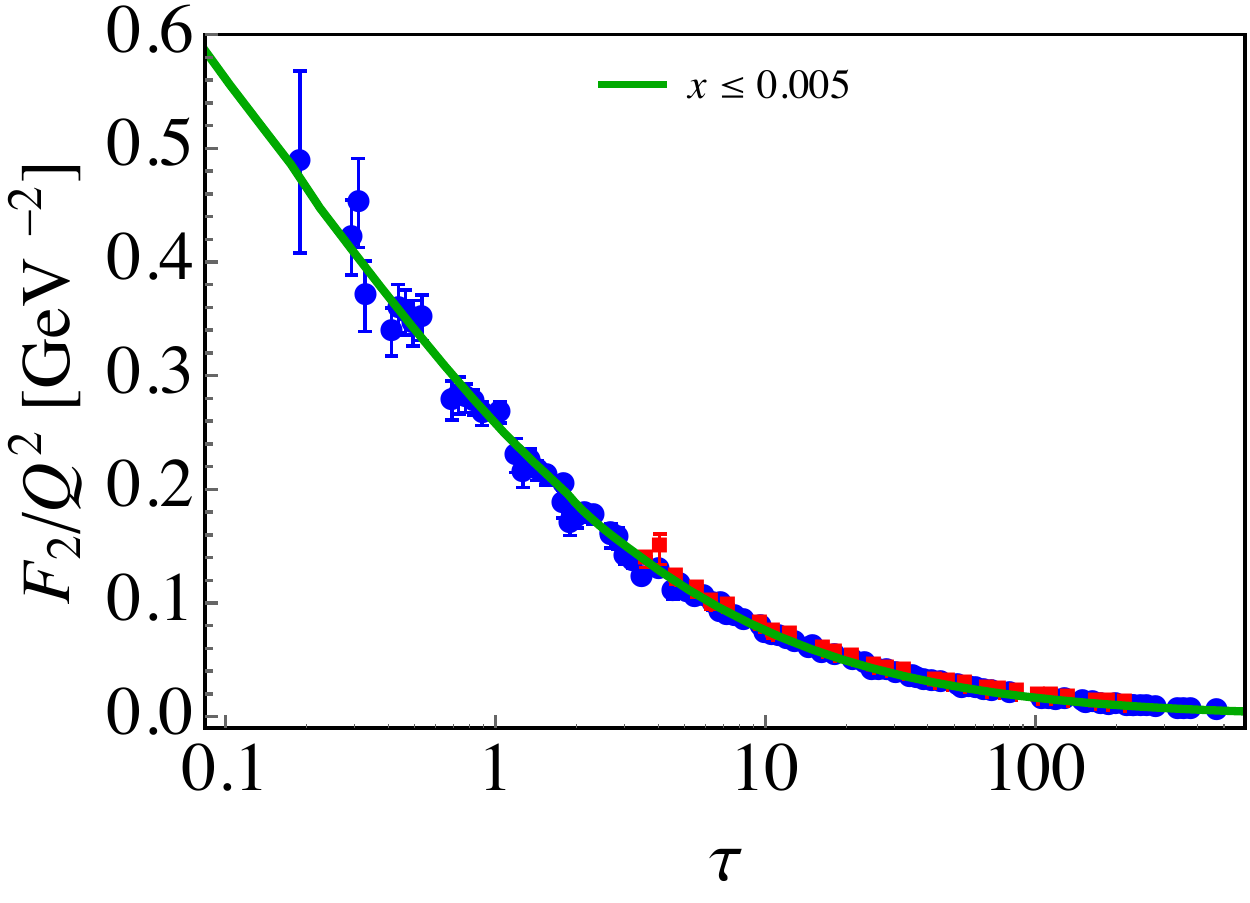}~\includegraphics[width=6.0cm]{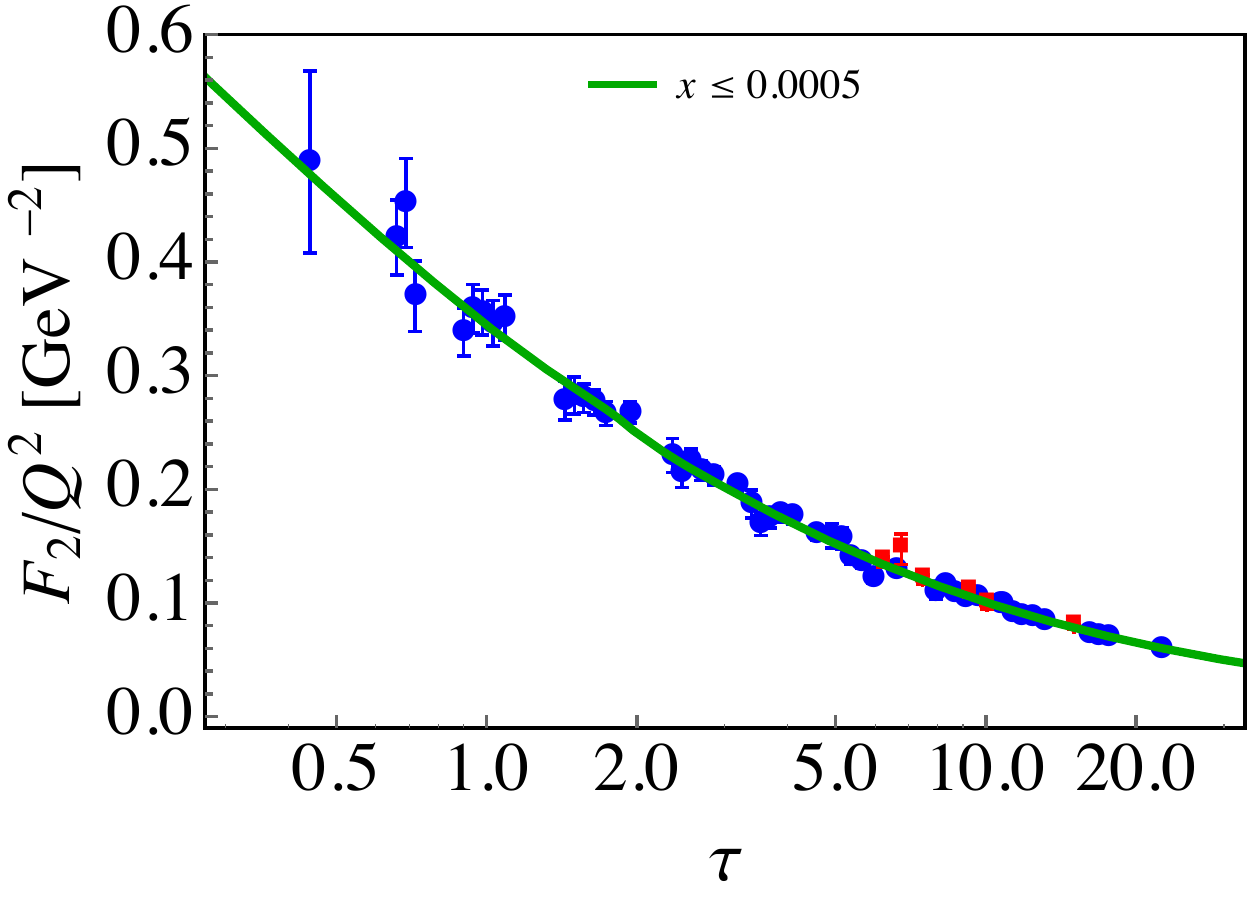}
\caption{H1
(blue circles) and ZEUS (red squares) data
\cite{Andreev:2013vha,Abramowicz:2014jak} for $F_{2}/Q^{2}$ 
plotted as a function of scaling variable $\tau$ for
different values of IIM model fit parameters given in Table~\ref{tab:IIM}.}%
\label{fig:F2_IIM}%
\end{figure}

However, there are two important differences between IIM and GBW
parametrizations. First of all, for small value of $u$ amplitude $A_{1}$
exhibits explicit violation of GS, since it depends both on $\tau$ and $x$.
For the purpose of the present work we will keep $x$ entering $A_{1}$ fixed at
the average value $x_{\text{ave}}$ calculated for the interval where the fit
is performed. The accuracy of this procedure is then checked by putting in
(\ref{sigmaIIM}) $x$ equal to the maximal and minimal value of $x$ in a given
interval. It will turn out that the structure functions are sensitive to this
variation of $x$ at the level of a few percent, however
the ratio $F_{\mathrm{L}}/F_{2}$ is almost independent. Next difference
appears due to the dependence of $A_{1}$ on $\lambda$. To solve this issue we
have adopted an iterative procedure, consisting in fixing $\lambda$ at some
initial value, then performing the fit and plugging in the fitted value of
$\lambda$ to $A_{1}$ and performing the fit again. Usually after four, five
steps satisfactory convergence has been  achieved. The results are
given in Table~\ref{tab:IIM}.

\begin{figure}[h]
\centering
\includegraphics[width=6.3cm]{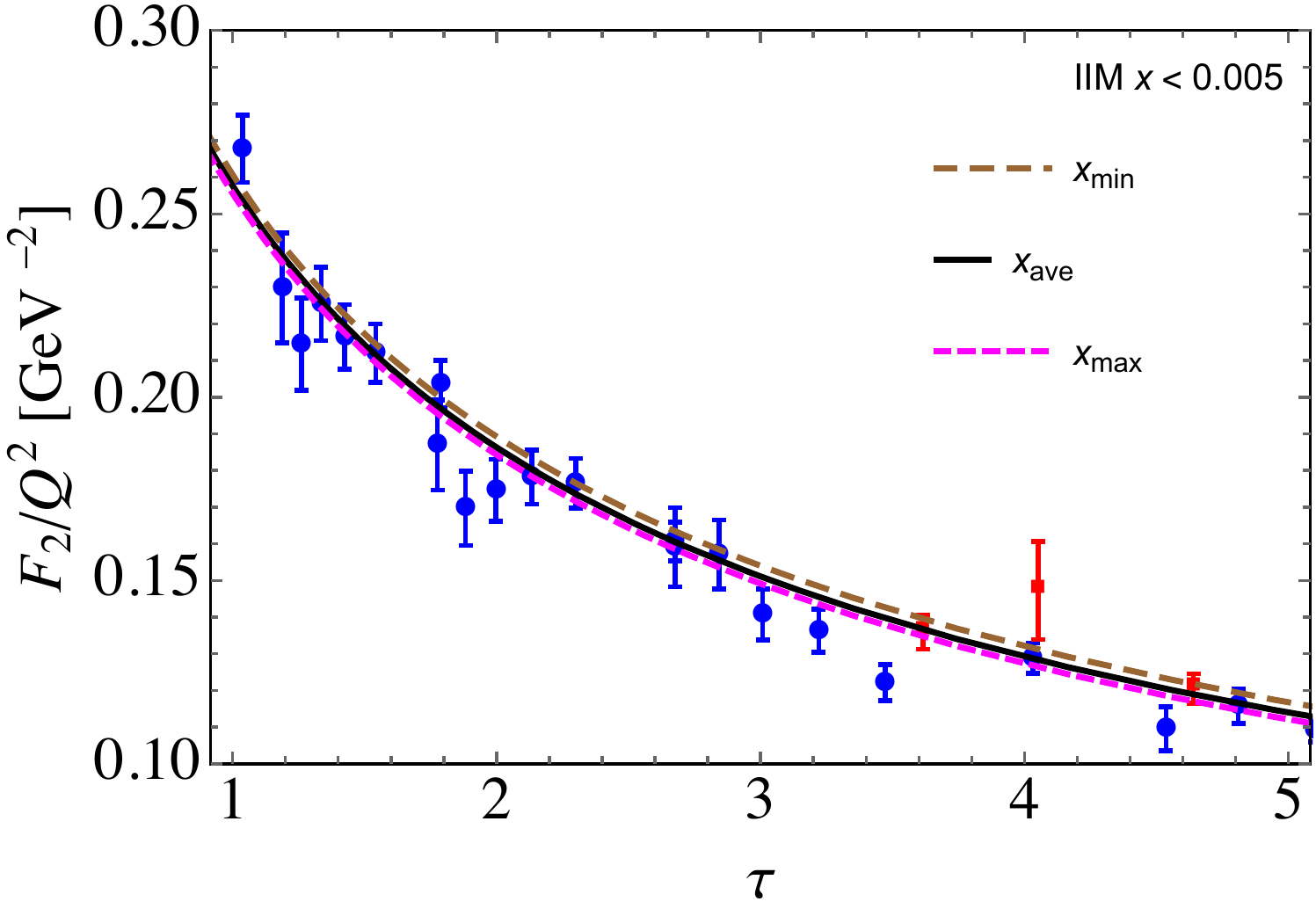}
\caption{$F_{2}(\tau)/Q^{2}$
for the parameters from the third row of
Table~\ref{tab:IIM} with $x$ in $A_{1}$ from Eq.~(\ref{A1A2_IIM})
replaced by $x_{\mathrm{min}}$ (upper
curve), $x_{\rm ave}$ (solid middle curve)  and $x_{\mathrm{max}}$ (lower curve). }%
\label{fig:F2_IIM_xdep}%
\end{figure}

By inspecting Table~\ref{tab:IIM} we see that the quality of fits is worse
than in the case of the GBW model. This is in contrast with the original fits
of Ref.~\cite{Iancu:2003ge} which, however, were performed over the data set
covering much lower $x$'s than in our present analysis. One should also note
that the errors of the old data sets are bigger than the ones of the present
data. This is also the reason why fit parameters are different than in our
case. For illustration purposes we have plotted in Fig.~\ref{fig:F2_IIM}
$F_{2}(\tau)/Q^{2}$ for the original IIM parameters and for three choices of
$x_{\mathrm{max}}$ from Table~\ref{tab:IIM}. 
Magnifying first plot in Fig.~\ref{fig:F2_IIM}
one could see that for $\tau> 10$
the original IIM curve missed the experimental points, which have rather small errors.

In order to check sensitivity of the IIM fits to the fact that $x$ dependendent
piece of $A_{1}$ amplitude (\ref{A1A2_IIM}) has been replaced by a constant
value $x_{\mathrm{ave}}$, we plot in Fig.~\ref{fig:F2_IIM_xdep} $F_{2}%
(\tau)/Q^{2}$ for the parameters from the third row 
of Table~\ref{tab:IIM} with $x$ in $A_{1}$ replaced by $x_{\mathrm{min}}$
(upper curve) and $x_{\mathrm{max}}$ (lower curve).
For better resolution the plot is restricted to $\tau< 5$. We can see that
theoretical uncertainty introduced by this procedure is in fact much smaller than
the experimental errors.

\begin{figure}[h]
\centering
\includegraphics[width=6.0cm]{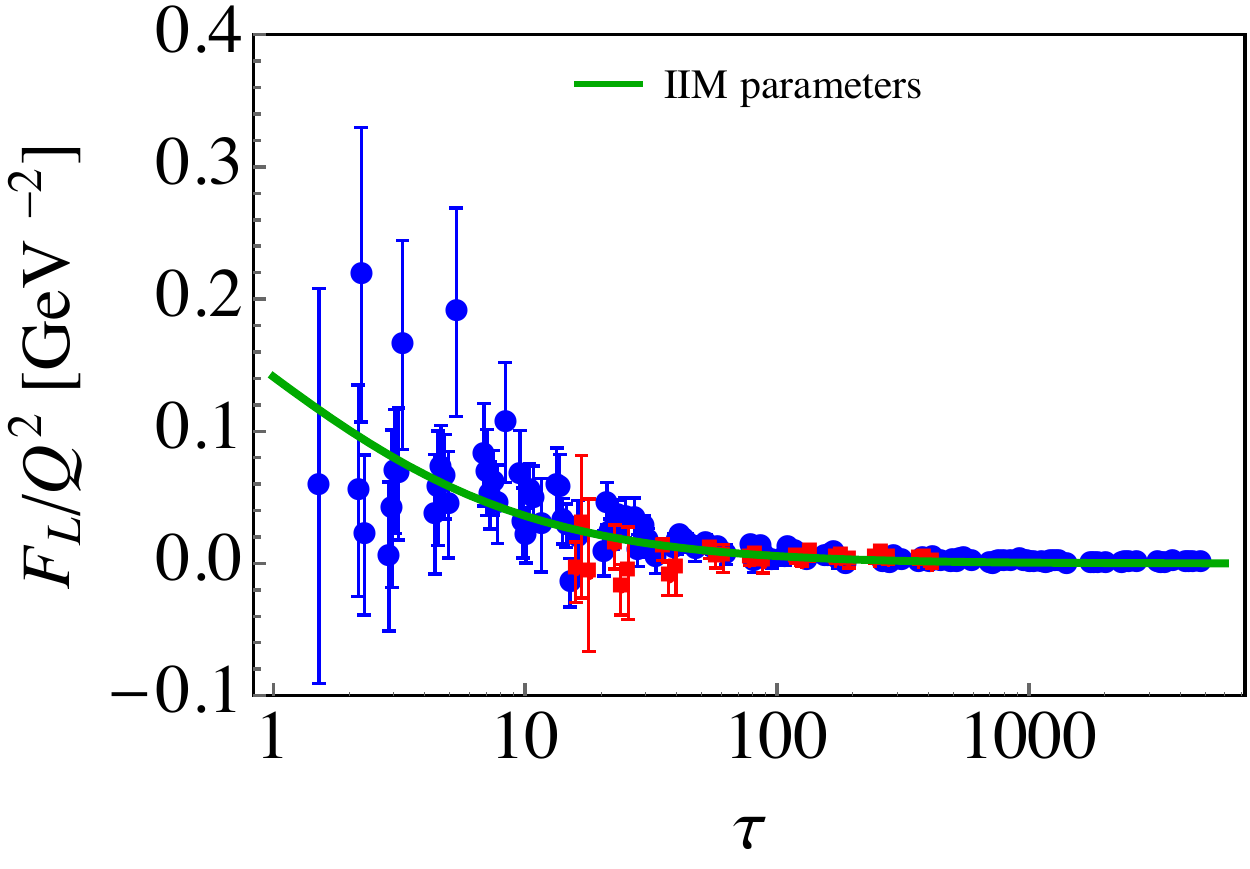}~\includegraphics[width=6.0cm]{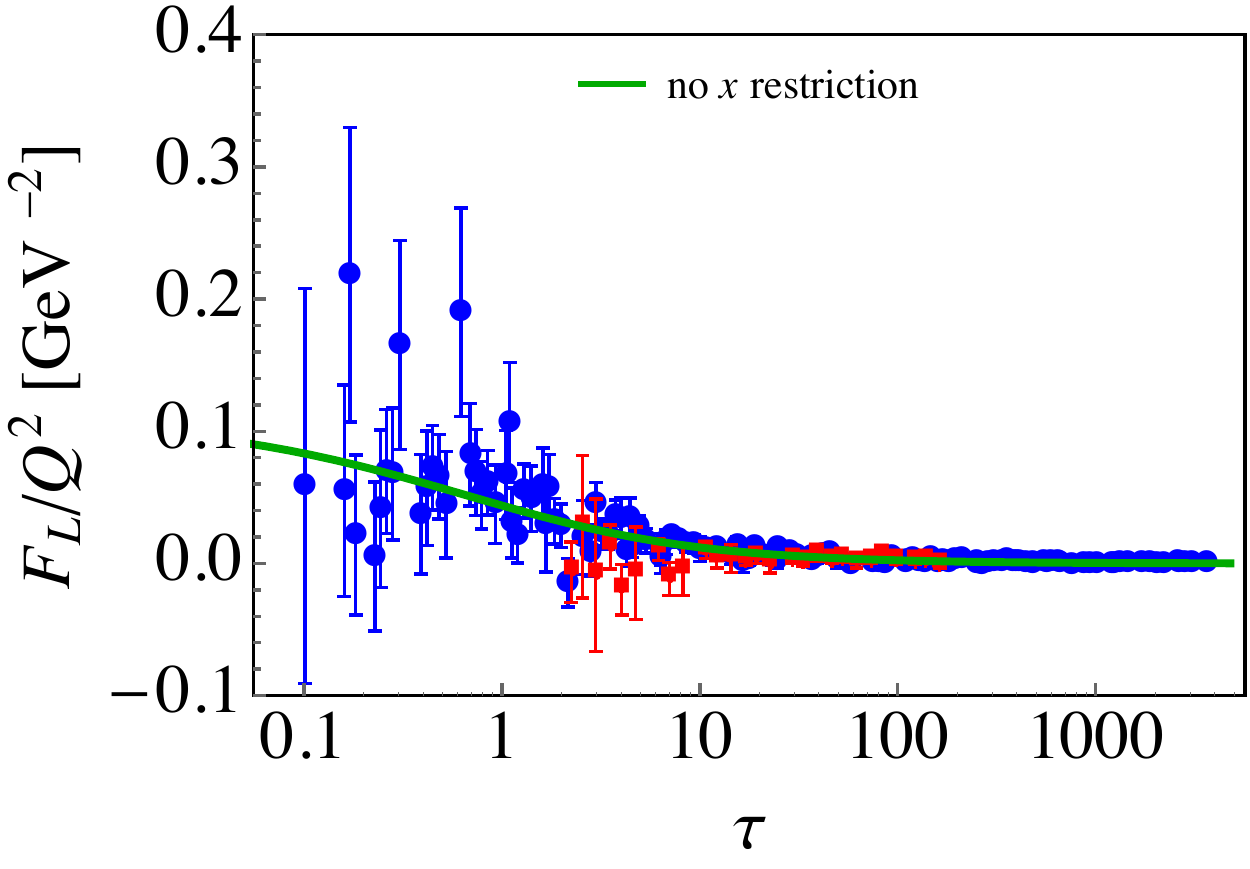}\\%
\includegraphics[width=6.0cm]{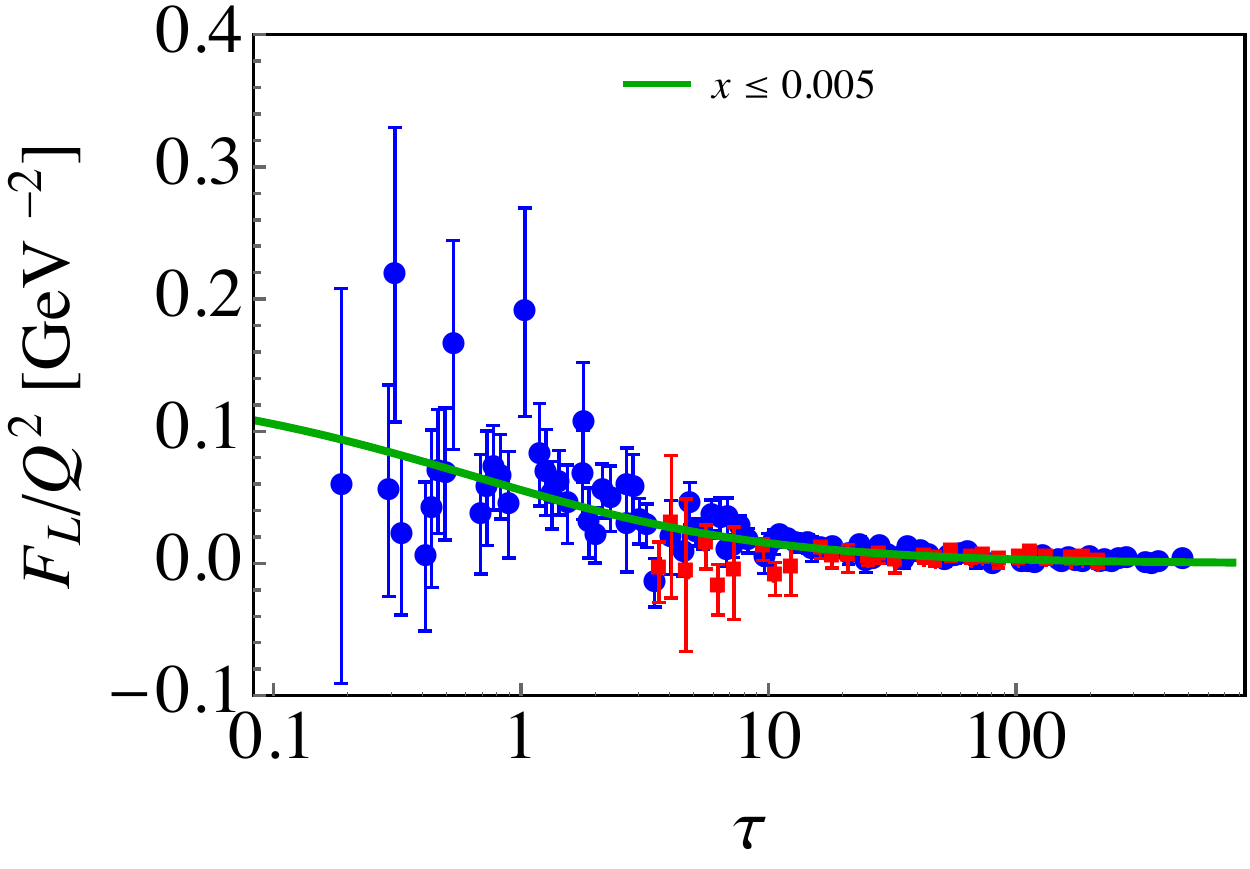}~\includegraphics[width=6.0cm]{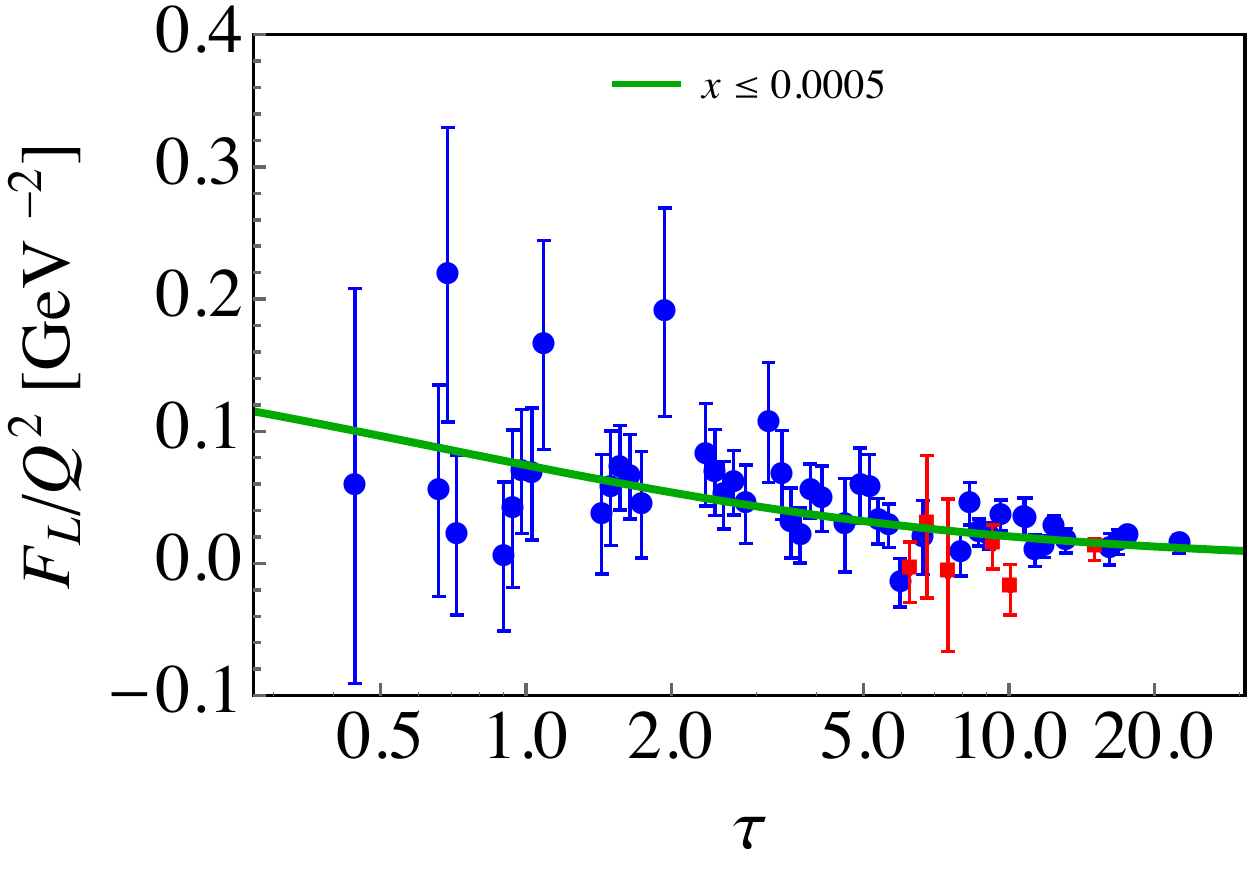}
\caption{H1
(blue circles) and ZEUS (red squares) data
\cite{Andreev:2013vha,Abramowicz:2014jak}  for $F_{\mathrm{L}%
}/Q^{2}$ plotted as a function of scaling variable $\tau$. Solid curves
correspond to the IIM model fits of Table~\ref{tab:IIM}. }%
\label{fig:FL_IIM}%
\end{figure}

Finally in Fig.~\ref{fig:FL_IIM} we plot $F_{\mathrm{L}}/Q^{2}$ as a
function of $\tau$ for three sets of parameters from 
Table~\ref{tab:IIM} and for the original set of parameters from
Ref.~\cite{Iancu:2003ge}. One can see that all curves describe the data
reasonably well due to the large error bars of $F_{\mathrm{L}}$.

\section{EMNS bound for dipole models}
\label{sec:modelsbound}

\begin{figure}[h]
\centering
\includegraphics[width=6.5cm]{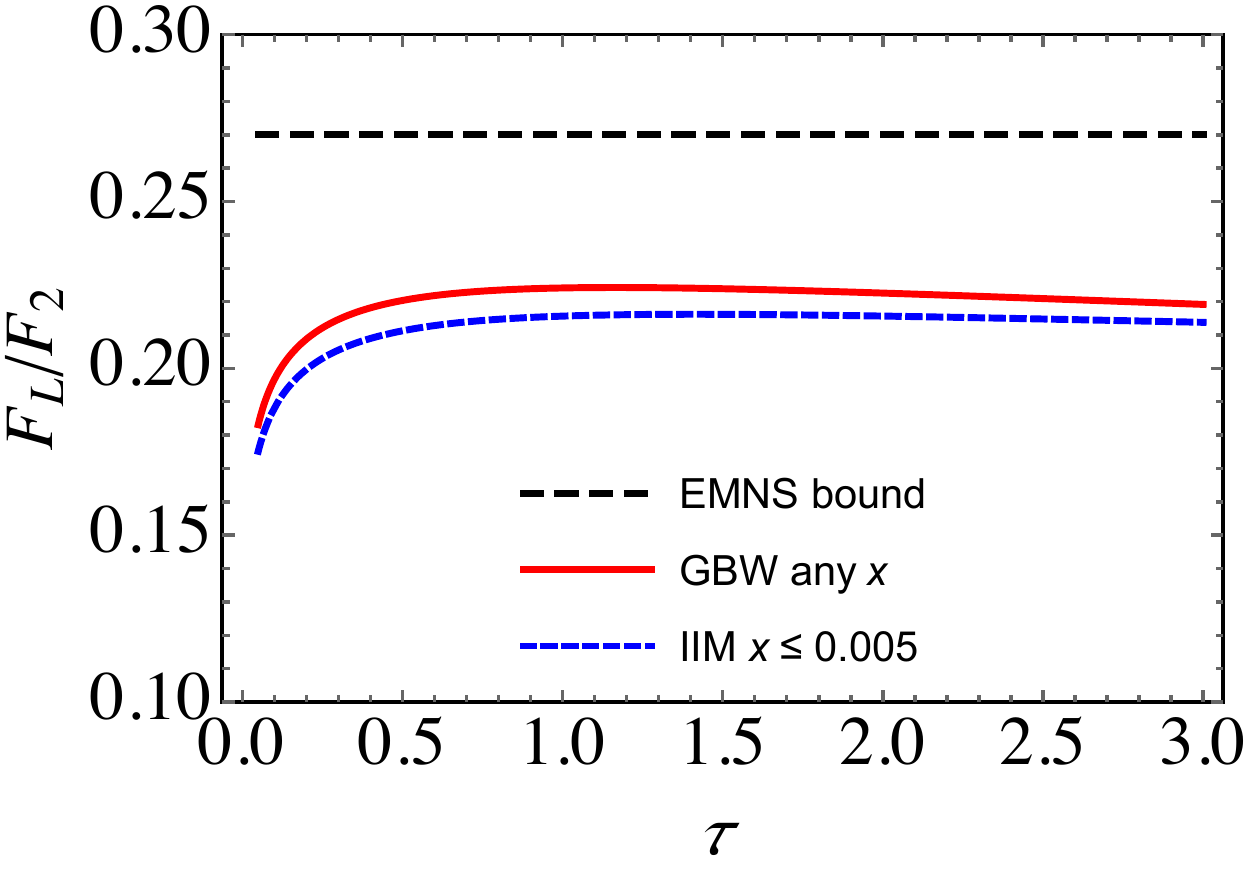}
\caption{Ratios $F_{\rm L}/F_2$ 
as functions of  $\tau$ for the GBW dipole model and for the IIM model with parameters corresponding to
the fit with $x<0.005$. The scale of the plot has been enlarged in order to make small differences
between the two curves visible.}%
\label{fig:model_bound}%
\end{figure}

Now we are able to compare the EMNS bound with  $F_{\rm L}/F_2$ ratio calculated in the dipole model
for realistic dipole-proton cross-sections $\sigma_{\rm dp}$
discussed in the previous Section. One should note that for the GBW dipole model
$F_{\rm L}/F_2$  does not depend on the values of $x_0$, $\lambda$ and $\sigma_0$. It is not the case
for the IIM dipole model, but we have checked explicitly that for all parametrizations  of Table~\ref{tab:IIM}
the differences in $F_{\rm L}/F_2$ are negligible. Therefore in Fig.~\ref{fig:model_bound} we plot ratios $F_{\rm L}/F_2$ 
as functions of  $\tau$ for the GBW dipole model and for the IIM model with parameters corresponding to
the fit with $x<0.005$. We see that ratios $F_{\rm L}/F_2$ are in fact almost model-independent. This is further
confirmed in Table~\ref{tab:ratio_max} where we collect the maximal value of $F_{\rm L}/F_2$ for the
GBW and IIM parametrizations. 

\begin{table}[t]
\centering
%%%table corrected%%%%%
\begin{tabular}
[c]{lcc}%
model & $\tau_{\text{max}}$  & $(F_{\rm L}/F_2)_{\rm max}$ \\\hline
GBW any& 1.165 & 0.224 \\
IIM all $x$ & 1.417 & 0.217 \\
IIM $x<0.01$ & 1.411 & 0.216 \\
IIM $x<0.005$ & 1.413 & 0.216 \\
IIM $x<0.0005$ & 1.418 & 0.217 \\\hline
\end{tabular}
\caption{Maxima of $F_{\rm L}/F_2$ for different fits to $F_2$.}%
\label{tab:ratio_max}%
\end{table}

\begin{figure}[h!]
\centering
\includegraphics[width=6.0cm]{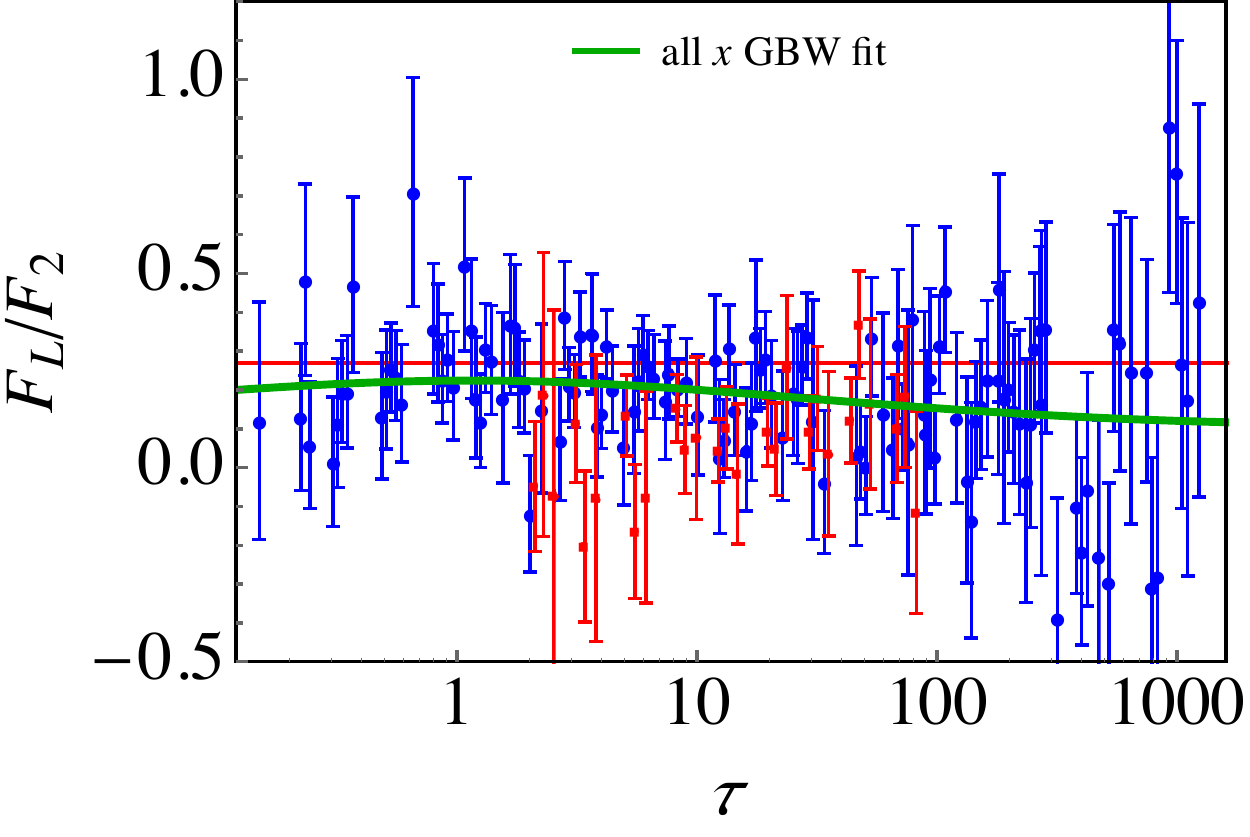}~\includegraphics[width=6.0cm]{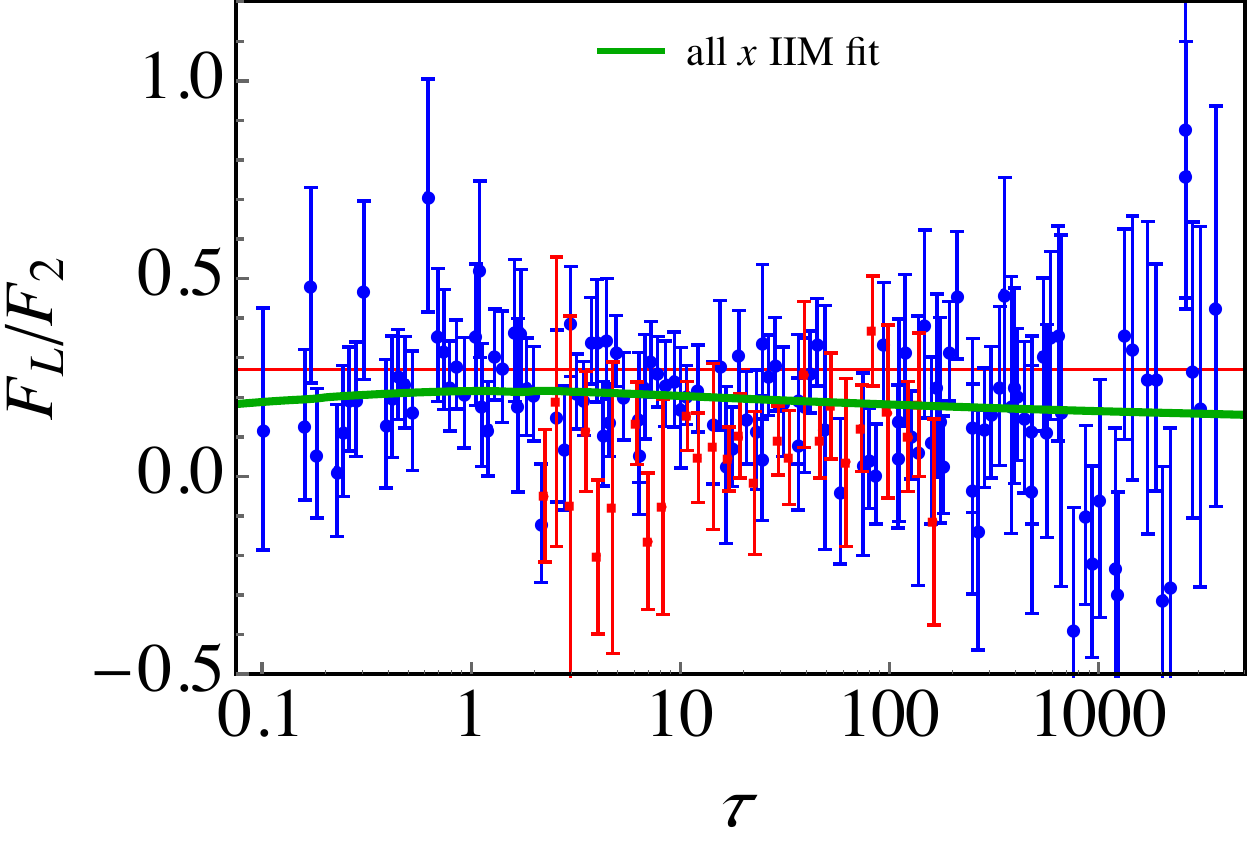}
\caption{Ratios $F_{\rm L}/F_2$ plotted as functions of scaling variable $\tau$ .
Straight line corresponds to the EMNS bound.
Solid line in the left panel corresponds to the unrestricted GBW fit of Tab.~\ref{tab:GBW} (first row), and in the
right panel to the unrestricted IIM fit of Tab.~\ref{tab:IIM} (first row). }
\label{fig:data_model_bound}%
\end{figure}

One should note that 
each curve in Fig.~\ref{fig:model_bound} corresponds to a different definition of scaling variable $\tau$, so one
cannot superimpose experimental data on that plot. This is done in Fig.~\ref{fig:data_model_bound} where we
plot $F_{\rm L}/F_2$ for the unrestricted fits of the GBW and IIM models corresponding to the first rows 
of Table~\ref{tab:GBW} and \ref{tab:IIM} respectively.
The errors of
the ratio have been calculated neglecting correlation between errors of $F_2$ and $F_{\rm L}$: 
\begin{equation}
\Delta\left( \frac{F_{\rm L}}{F_2} \right)=\frac{F_{\rm L}}{F_2}
\sqrt{\left( \frac{\Delta F_{\rm L}}{F_L} \right)^2 + \left( \frac{\Delta F_{\rm 2}}{F_2} \right)^2}
\label{eq:erratio}
\end{equation}
This procedure overestimates the errors, however, given the fact that $F_L<0.27 F_{\rm 2}$ and that
experimentally absolute errors of $F_{\rm L}$ are 2 -- 10 times larger than $\Delta F_2$, the error 
of the ratio (\ref{eq:erratio}) is determined to very high precision by $\Delta F_{\rm L}$ alone.
 We shall come back to this point later.

We can see from Figs.~\ref{fig:model_bound} and \ref{fig:data_model_bound} that for realistic $\sigma_{\rm dp}$
theoretical predictions lie below the EMNS bound. Indeed, we see that the maximum of $F_{\rm L}/F_2$ is of the order
$0.216 - 0.224$ and only slightly varies from fit to fit. This is illustrated in Table~\ref{tab:ratio_max}. A question arises whether
data points -- which for some values of $\tau$ exceed the EMNS bound -- are indeed, as suggested by the authors of 
Ref.~\cite{Ewerz:2012az}, saturating bound (\ref{Nbound}), being as a consequence incompatible with 
the dipole model.  To this end we have simply calculated $\chi^2$ of $F_{\rm L}/F_2$ for the unrestricted GBW fit 
and obtained very good result: 0.7425. To check whether this value is affected by the fact that we have not
taken into account correlations between errors of $F_{\rm L}$ and $F_2$, we have recalculated $\chi^2$
neglecting $F_2$ errors, which gives $\chi^2$ that changes by less than 1\%.  Therefore indeed, as already
mentioned above,  $\chi^2$
value is totally driven by the large errors of $F_{\rm L}$.

\begin{figure}[h!]
\centering
\includegraphics[width=6.0cm]{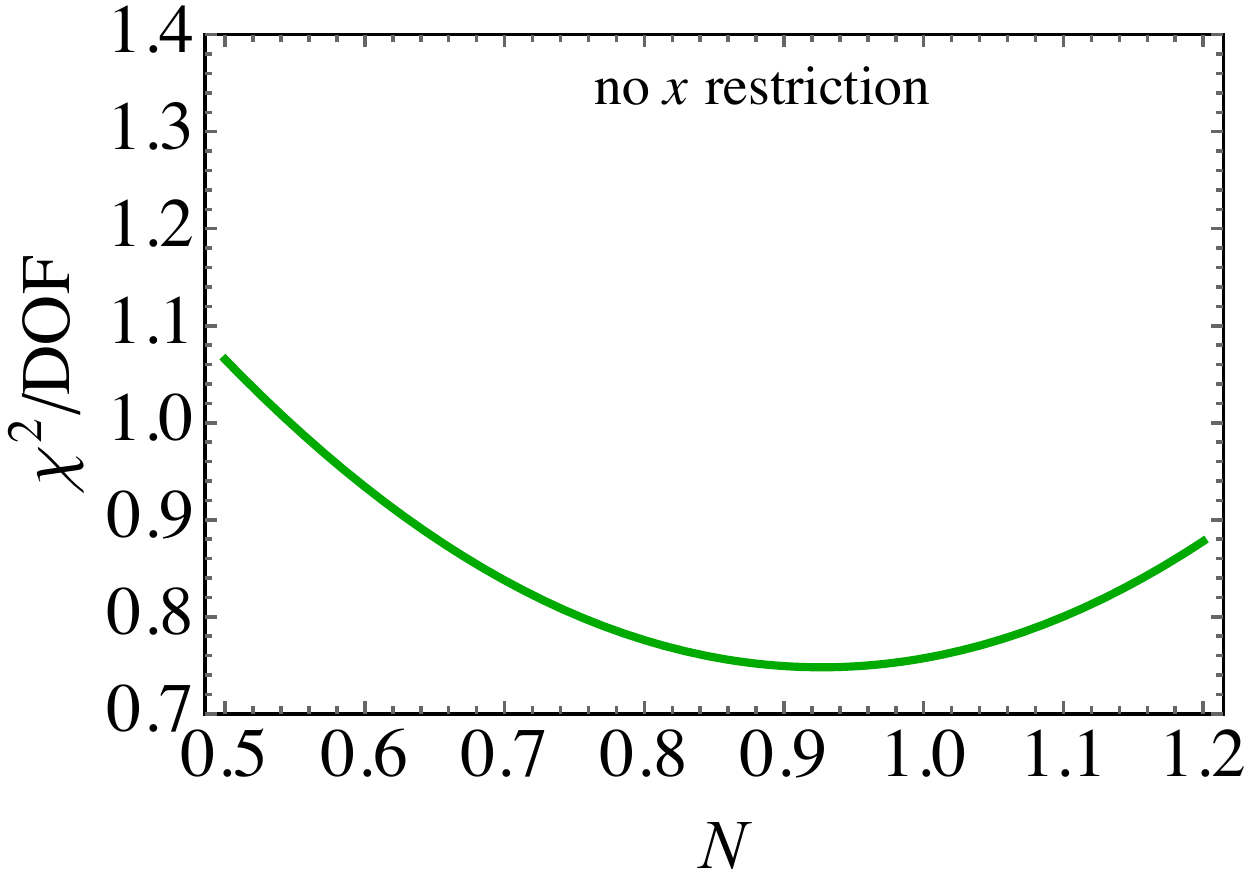}~\includegraphics[width=6.0cm]{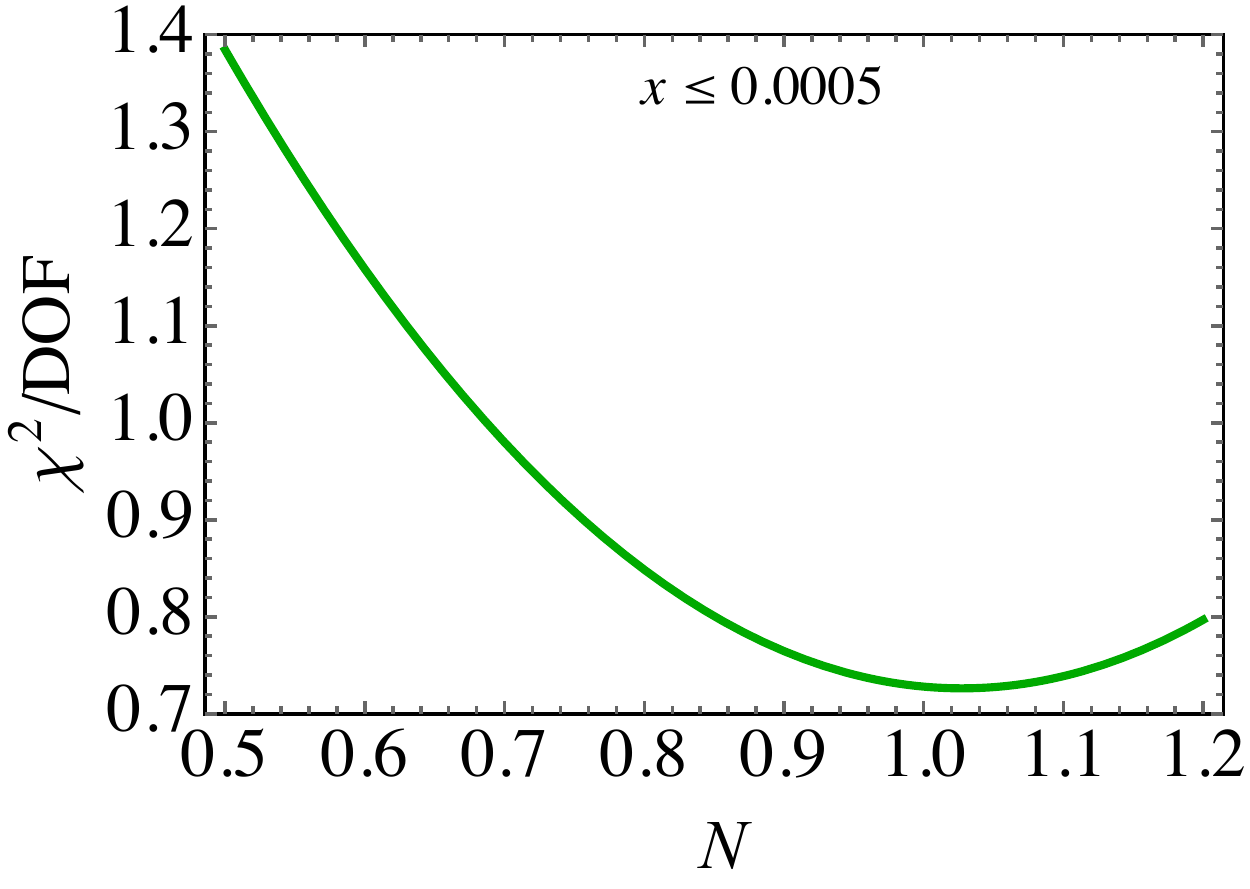}
\caption{Change of $\chi^2$ of ${\cal N} \times F_{\rm L}/F_2$ as a function of  ${\cal N}$ for the GBW
fits from Table~\ref{tab:GBW} for unrestricted $x$ (left) and $x<0.0005$ (right).}
\label{fig:chi2_4_N}%
\end{figure}

It is interesting to check whether an overall shift of the dipole model prediction for $F_{\rm L}/F_2$
would improve  agreement  with the data. In this way we shall have a quantitative measure 
of the quality of the dipole model prediction
for the ratio $F_{\rm L}/F_2$ and also an indication how much room is there for the higher order corrections 
that we are going to discuss in Sect.~\ref{sec:HFS}.
To this end we use the GBW fits allowing
for arbitrary normalization of the ratio
\begin{equation}
\frac{F_{\rm L}}{F_2} \rightarrow {\cal N} \frac{F_{\rm L}}{F_2}
\end{equation}
and calculate  $\chi^2$/d.o.f. (assuming one degree of freedom, namely  ${\cal N}$)
as a function of  ${\cal N}$. This is illustrated in Fig.~\ref{fig:chi2_4_N} for two GBW fits of Table~\ref{tab:GBW}:
unrestricted $x$ and $x<0.0005$. We see that depending on the fit the data prefers ${\cal N}$ sligthly smaller (unrestricted
$x$) or sligthly larger ($x<0.0005$) than 1.
These are negligible changes and therefore at this stage we conclude
that the data for the ratio $F_{\rm L}/F_2$ is compatible with the dipole model. One should perhaps
remind here again, that this ratio is quite stable, even for fits to $F_2$ that are visibly different.

One can also see that lowering ${\cal N}$ by $~20\%$, which would be requried by taking into account
charm mass effects ({\em c.f.} discussion at the end of Sect.~\ref{sec:GSandbound}) increases $\chi^2$,
but only by a small ammount. Of course the detialed study of charm mass effects would require to go 
beyond GS used in this  analysis -- which in turn would result not only in the change of $\cal N$, but
also in the change of the shape of $F_{\rm L}/F_2$ --
but no drastic difference to the present analysis should be expected.

\section{EMNS bound and higher Fock states}
\label{sec:HFS}

In the dipole model the virtual photon dissociates into a $\bar{q}q$ pair
which subsequently interacts with the proton target. However, it is clear that
higher Fock states have to contribute as well, similarly to the diffractive DIS
where the next Fock component, namely the $\bar{q}qg$ state, is dominant at small
$\beta$ \cite{GolecBiernat:1999qd}. Full calculations of the $\bar{q}qg$ component
of the photon wave function have
recently appeared in the literature \cite{Balitsky:2010ze,Beuf:2011xd,Boussarie:2014lxa},
however they have not been yet applied to the phenomenological analysis of DIS.

Structure functions in the dipole model are given as an
expansion%
\begin{equation}
F_{2,L}(x,Q^{2})=F_{2,L}^{(\bar{q}q)}(x,Q^{2})+F_{2,L}^{(\bar{q}qg)}%
(x,Q^{2})+\ldots\,.
\label{expansion}
\end{equation}
The EMNS bound (\ref{Nbound}) derived in Sect.~\ref{sec:GSandbound} is in fact valid
only for the first component of (\ref{expansion})%
\begin{equation}
G=\frac{F_{\mathrm{L}}^{(\bar{q}q)}(x,Q^{2})}{F_{2}^{(\bar{q}q)}(x,Q^{2}%
)}<g_{\rm max}=0.27.
\label{Nbound1}
\end{equation}
One should note, however, that at this order of perturbative expansion the loop corrections
to the leading order Fock component may change the value of the bound (\ref{Nbound1}).
Similarly for the next Fock component we would have:
\begin{equation}
H=\frac{F_{\mathrm{L}}^{(\bar{q}qg)}(x,Q^{2})}{F_{2}^{(\bar{q}qg)}(x,Q^{2})}<h_{\rm max}
\end{equation}
where%
\begin{equation}
0\leq h_{\rm max}\leq1.
\end{equation}
Up to this order one can derive the {\em modified} EMNS bound:%
\begin{align}
\frac{F_{\mathrm{L}}(x,Q^{2})}{F_{2}(x,Q^{2})}  & =\frac{F_{\mathrm{L}}%
^{(\bar{q}q)}(x,Q^{2})+F_{\mathrm{L}}^{(\bar{q}qg)}(x,Q^{2})}{F_{2}^{(\bar
{q}q)}(x,Q^{2})+F_{2}^{(\bar{q}qg)}(x,Q^{2})}\nonumber\\
& <\frac{g_{\rm max}\,F_{2}^{(\bar{q}q)}(x,Q^{2})+h_{\rm max}\,F_{2}^{(\bar{q}qg)}(x,Q^{2})}%
{F_{2}^{(\bar{q}q)}(x,Q^{2})+F_{2}^{(\bar{q}qg)}(x,Q^{2})}\nonumber\\
& =g_{\rm max} \frac{1+\delta\,\varepsilon(x,Q^{2})}{1+\varepsilon(x,Q^{2})}%
\end{align}
where%
\[
\delta=\frac{h_{\rm max}}{g_{\rm max}},\qquad0\leq\delta\leq\frac{1}{g_{\rm max}}\simeq3.7
\]
and%
\begin{equation}
\varepsilon(x,Q^{2})=\frac{F_{2}^{(\bar{q}qg)}(x,Q^{2})}{F_{2}^{(\bar{q}%
q)}(x,Q^{2})}.
\end{equation}
Since the dipole model with $\bar{q}q$ component only describes $F_{2}$
rather well over the wide kinematical range, we do not expect $\varepsilon$ to
be large. For the purpose of the present analysis we assume that $\varepsilon$
does not exceed 20\%. 

\begin{figure}[h!]
\centering
\includegraphics[height=7cm]{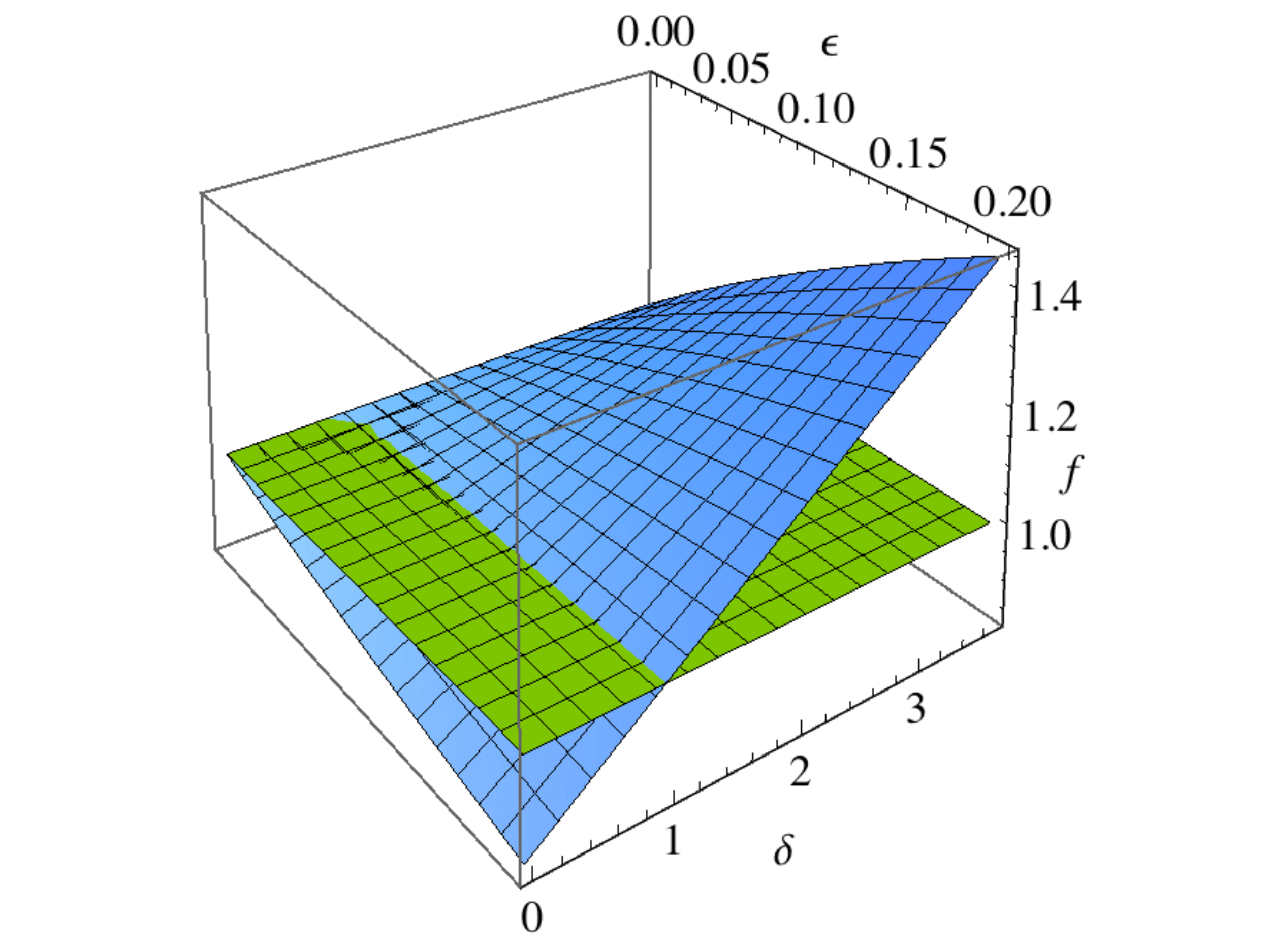}\caption{Modification factor
$f$ of Eq.~(\ref{fmod}) as a function of $\epsilon$ and $\delta$. Light green
plane corresponds to $f=1$.}%
\label{fig:modification}%
\end{figure}

In order to change  the value of the EMNS bound we need 
the {\em modification factor}%
\begin{equation}
f=\frac{1+\varepsilon\,\delta}{1+\varepsilon}\label{fmod}%
\end{equation}
to be significantly different from 1. In Fig.~\ref{fig:modification} we plot $f(\varepsilon,\delta)$ for $0\leq
\varepsilon\leq0.2$ and $0\leq\delta\leq1/g_{\rm max}.$ We see from Fig.~\ref{fig:modification}
and from Eq.~(\ref{fmod}) that $f=1$ for $\delta=1$ and that $f$ gets smaller than 1 if $\delta<1$ and
$f>1$ for $\delta>1$, and that the difference $|{f-1}|$
 is growing with $\varepsilon$.\ From Fig.~\ref{fig:modification} one may conclude that
the modification of the EMNS bound by more than $\pm$20\% might be rather difficult.
Most probably higher Fock components would modify (\ref{Nbound}) by less than 10\%,
but to quantify this statement one needs to calculate explicitly  $h_{\rm max}$
which is beyond the scope of the present paper. One should note at this point that in this
case an inclusion of mass corrections due to the charm and possibly bottom quarks should
be included, as these corrections  would be of the same order or even larger than a
contribution from the higher Fock components. 

To conclude this Section let us only remark
that $\varepsilon$ depends on kinematical variables, so one cannot exclude {\em a'priori}
a situation that there exists a kinematical corner where a correction to  (\ref{Nbound})
is of importance.

\section{Conclusions}
\label{sec:conclusions}

Dipole model offers effective and intuitive description of deep inelastic scattering 
which goes beyond leading twist approximation. In this paper
we have used different variants of the dipole model, {\em i.e.} two different forms of
photon-proton cross-section fitted to $F_2$ data over different ranges of Bjorken $x$'s. We
have decided to perform our new fits in the kinematical range where both $F_2$ and $F_{\rm L}$
have been measured. This kinematical region does not extend to very small $x$'s as it is in the
case of the recently published combined HERA data \cite{HERAcombined} and therefore
fitted parameters are different than the ones of global fits. For the same reason the IIM model
\cite{Iancu:2003ge} that has been specifically devised for low $x$ region gives larger $\chi^2$ than the
simplest version of the the GBW model \cite{GolecBiernat:1998js}, which quite satisfactorily describes
$F_2$. 

In order to fit model parameters we have used the property of geometrical scaling, which boils
down to the fact that data points of the same $Q^2$ but different $x$'s disperse when plotted in terms 
of scaling variable $\tau$ and fall on one line (compare Figs.~\ref{fig:F2L_data} and {\em e.g.} \ref{fig:F2_GBW}).
We have found that GS is present both in the case of $F_2$ and $F_{\rm L}$ as well.
One should, however, take this property with care for the present set of data, since for each value of $Q^2$
only a few points of different Bjorken $x$'s have been measured. This $x - Q^2$ correlation of the DIS
data, particularly pronounced in the present case of $F_{\rm L}$, is a common obstacle in deriving firm
conclusions on the quality of GS. In this context it is also worth mentioning that there exist different sources
of GS. The first one is related to the genuine property of the initial state of the proton target described
theoretically by the Color Glass Condensate formalism \cite{CGC} and nonlinearities in the
parton evolution \cite{MunPesch}, which are of importance at small values of Bjorken $x$. 
The other one is a property of  the linear parton evolution equations \cite{Forte}, which build GS
at larger $Q^2$ and at not necessarily very small $x$'s.

We have next compared dipole model predictions -- with parameters fixed by fits to $F_2$ -- with
$F_{\rm L}$ data. Here, due to  large experimental errors,  agreement is quite good. Next
we have studied ratio of the structure functions $G=F_{\rm L}/F_2$ for which in the dipole model there exists 
a strict bound $g_{\rm max}=0.27$ (\ref{bound}) derived in Ref.~\cite{Ewerz:2012az}. 
We have rederived (\ref{bound}) with the help of geometrical scaling. We have also shown that in the dipole
models discussed above $g_{\rm max}^{\rm dp} \approx 0.216 - 0.224$ which is approximately 18.5\% below
the EMNS bound. Different fits give very similar ratio $g$,  which is only residually dependent
on the values of fit parameters (this concerns only the IIM dipole model, since ratio $F_{\rm L}/F_2$ is parameter
independent in the GBW case). Comparing $G=F_{\rm L}/F_2$ with 
the data we have established that the GBW model  reproduces $G$ with high precision. We do not see
any tension between the data and the dipole model as far as ratio $G$ is concerned, even if charm mass effects,
which lower the EMNS bound, are taken into account.

Dipole model and the EMNS bound discussed so far  rely on the first Fock component of the photon wave function.
Including higher Fock components, like a $\bar{q}qg$ state, might in principle change theoretical prediction
for $g_{\rm max}$. For this to happen, longitudinal part of the $\bar{q}qg$ state compared to the transverse one
has to be significantly different than in the $\bar{q}q$ case. 

\section*{Acknowledgments}
The authors want to thank Aharon Levy  for bringing to their attention new data on $F_{\rm L}$ and for 
an access to this data prior to publication. M.P. thanks Otto Nachtmann for a conversation
that started this project and for  remarks on the final version of the manuscript.
We thank Guillaume Beuf for bringing to our attention Refs.~\cite{Beuf:2011xd,Boussarie:2014lxa}.
The research of M.P. has been supported by the Polish NCN grants 2011/01/B/ST2/00492
and 2014/13/B/ST2/02486.


\newpage

\begin{thebibliography}{99}                                                                                               %
\bibitem {Andreev:2013vha}V.~Andreev \textit{et al.} [H1 Collaboration],
%``Measurement of inclusive $e p$ cross sections at high $Q^2$ at $\sqrt s =$ 225 and 252 GeV and of the longitudinal proton structure function $F_{\rm L}$ at HERA,''
Eur.\ Phys.\ J.\ C \textbf{74} (2014) 2814 [arXiv:1312.4821 [hep-ex]].
%%CITATION = ARXIV:1312.4821;%%
%4 citations counted in INSPIRE as of 04 Aug 2014
%\cite{Abramowicz:2014jak}


\bibitem {Abramowicz:2014jak}H.~Abramowicz \textit{et al.} [ZEUS
Collaboration],
%``Deep inelastic cross-section measurements at large y with the ZEUS detector at HERA,''
arXiv:1404.6376 [hep-ex].
%%CITATION = ARXIV:1404.6376;%%
%1 citations counted in INSPIRE as of 04 Aug 2014


%\cite{Aaron:2008ad}


\bibitem {Aaron:2008ad} F.~D.~Aaron \textit{et al.} [H1 Collaboration],
%``Measurement of the Proton Structure Function F(L)(x, Q**2) at Low x,''
Phys.\ Lett.\ B \textbf{665} (2008) 139  [arXiv:0805.2809 [hep-ex]] and
%%CITATION = ARXIV:0805.2809;%%
%107 citations counted in INSPIRE as of 30 Nov 2014\
%\cite{Collaboration:2010ry}
%\bibitem{Collaboration:2010ry}
%F.~D.~Aaron {\it et al.}  [ H1 Collaboration],
%``Measurement of the Inclusive e{\pm}p Scattering Cross Section at High Inelasticity y
%and of the Structure Function $F_L$,''
Eur.\ Phys.\ J.\ C \textbf{71} (2011) 1579  [arXiv:1012.4355 [hep-ex]].
%%CITATION = ARXIV:1012.4355;%%
%68 citations counted in INSPIRE as of 30 Nov 2014


%\cite{Levy:2014tna}


\bibitem {Levy:2014tna}A.~Levy, talk at QCD Moriond 2014,
%Recent HERA results on proton structure,''
arXiv:1405.3753 [hep-ex].
%%CITATION = ARXIV:1405.3753;%%
%\cite{Andreev:2013vha}


%\cite{Callan:1969uq}


\bibitem {Callan:1969uq} C.~G.~Callan, Jr. and D.~J.~Gross,
%``High-energy electroproduction and the constitution of the electric current,''
Phys.\ Rev.\ Lett.\ \textbf{22} (1969) 156.
%%CITATION = PRLTA,22,156;%%
%818 citations counted in INSPIRE as of 30 Nov 2014

%\cite{Ewerz:2006an}


\bibitem {Ewerz:2006an} C.~Ewerz and O.~Nachtmann,
%``Bounds on Ratios of DIS Structure Functions from the Color Dipole Picture,''
Phys.\ Lett.\ B \textbf{648} (2007) 279  [hep-ph/0611076].
%%CITATION = HEP-PH/0611076;%%
%7 citations counted in INSPIRE as of 30 Nov 2014


%\cite{Ewerz:2007md}
\bibitem{Ewerz:2007md}
  C.~Ewerz, A.~von Manteuffel and O.~Nachtmann,
  %``On the Range of Validity of the Dipole Picture,''
  Phys.\ Rev.\ D {\bf 77} (2008) 074022
  [arXiv:0708.3455 [hep-ph]].
  %%CITATION = ARXIV:0708.3455;%%
  %9 citations counted in INSPIRE as of 14 Feb 2015

%\cite{Ewerz:2012az}


\bibitem {Ewerz:2012az} C.~Ewerz, A.~von Manteuffel, O.~Nachtmann and
A.~Sch{\"o}ning,
%``The New F_L Measurement from HERA and the Dipole Model,''
Phys.\ Lett.\ B \textbf{720} (2013) 181  [arXiv:1201.6296 [hep-ph]].
%%CITATION = ARXIV:1201.6296;%%
%3 citations counted in INSPIRE as of 30 Nov 2014

%\cite{GolecBiernat:1998js}


\bibitem {GolecBiernat:1998js} K.J.~Golec-Biernat and M.~W{\"u}sthoff,
%``Saturation effects in deep inelastic scattering at low Q**2 and its implications on diffraction,''
Phys.\ Rev.\ D \textbf{59} (1998) 014017  [hep-ph/9807513].
%%CITATION = HEP-PH/9807513;%%
%905 citations counted in INSPIRE as of 30 Nov 2014


%\cite{GolecBiernat:1999qd}


\bibitem {GolecBiernat:1999qd} K.J.~Golec-Biernat and M.~W{\"u}sthoff,
%``Saturation in diffractive deep inelastic scattering,''
Phys.\ Rev.\ D \textbf{60} (1999) 114023  [hep-ph/9903358].
%%CITATION = HEP-PH/9903358;%%
%680 citations counted in INSPIRE as of 30 Nov 2014



%\cite{Ewerz:2006vd}
\bibitem{Ewerz:2006vd}
  C.~Ewerz and O.~Nachtmann,
  %``Towards a nonperturbative foundation of the dipole picture. II. High energy limit,''
  Annals Phys.\  {\bf 322} (2007) 1670
  [hep-ph/0604087].
  %%CITATION = HEP-PH/0604087;%%
  %25 citations counted in INSPIRE as of 14 Feb 2015


\bibitem {Stasto:2000er} A.M.~Stasto, K.J.~Golec-Biernat, J.~Kwiecinski,
%\emph{Geometric scaling for the total $\gamma^{*}p$ cross-section in the low x
%region},
{Phys.\ Rev.\ Lett.}\ \textbf{86} (2001) 596.
%[arXiv:hep-ph/0007192].
%%CITATION = PRLTA,86,596;%%


%\cite{Praszalowicz:2012zh}


\bibitem {Praszalowicz:2012zh} M.~Praszalowicz and T.~Stebel,
%``Quantitative Study of Geometrical Scaling in Deep Inelastic Scattering at HERA,''
JHEP \textbf{1303} (2013) 090  [arXiv:1211.5305 [hep-ph]].
%%CITATION = ARXIV:1211.5305;%%
%10 citations counted in INSPIRE as of 30 Nov 2014

\bibitem{Balitsky:2010ze}
  I.~Balitsky and G.~A.~Chirilli,
  %``Photon impact factor in the next-to-leading order,''
  Phys.\ Rev.\ D {\bf 83} (2011) 031502
  [arXiv:1009.4729 [hep-ph]] and
  %``Photon impact factor and $k_T$-factorization for DIS in the next-to-leading order,''
  Phys.\ Rev.\ D {\bf 87} (2013) 1,  014013
  [arXiv:1207.3844 [hep-ph]].
  %%CITATION = ARXIV:1207.3844;%%
  %25 citations counted in INSPIRE as of 17 Feb 2015
  %%CITATION = ARXIV:1009.4729;%%
  %42 citations counted in INSPIRE as of 17 Feb 2015

%\cite{Beuf:2011xd}
\bibitem{Beuf:2011xd}
  G.~Beuf,
  %``NLO corrections for the dipole factorization of DIS structure functions at low x,''
  Phys.\ Rev.\ D {\bf 85} (2012) 034039
  [arXiv:1112.4501 [hep-ph]].
  %%CITATION = ARXIV:1112.4501;%%
  %15 citations counted in INSPIRE as of 17 Feb 2015
  
%\cite{Boussarie:2014lxa}
\bibitem{Boussarie:2014lxa}
  R.~Boussarie, A.~V.~Grabovsky, L.~Szymanowski and S.~Wallon,
  %``Impact factor for high-energy two and three jets diffractive production,''
  JHEP {\bf 1409} (2014) 026
  [arXiv:1405.7676 [hep-ph]].
  %%CITATION = ARXIV:1405.7676;%%


\bibitem {empty}See {\em e.g.} Chapter 9 in V.~Baronne and E.~Predazzi,  {\em 
High-Energy Particle Diffraction}, Springer 2002.


%\cite{Iancu:2003ge}


\bibitem {Iancu:2003ge} E.~Iancu, K.~Itakura and S.~Munier,
%``Saturation and BFKL dynamics in the HERA data at small x,''
Phys.\ Lett.\ B \textbf{590} (2004) 199  [hep-ph/0310338].
%%CITATION = HEP-PH/0310338;%%
%317 citations counted in INSPIRE as of 02 Dec 2014


\bibitem {HERAcombined}F.~D.~Aaron \textit{et al.} [H1 and ZEUS
Collaboration],
%``Combined Measurement and QCD Analysis of the Inclusive ep Scattering Cross
%Sections at HERA,''
JHEP \textbf{1001} (2010) 109 [arXiv:0911.0884 [hep-ex]].
%%CITATION = JHEPA,1001,109;%%



\bibitem{CGC}
%\bibitem {MLV} 
L.~D.~McLerran, and R.~Venugopalan,
%``Computing quark and gluon distribution functions for very large nuclei,''
Phys.\ Rev.\ \textbf{D49}, 2233 (1994), \ Phys. Rev. \textbf{D49}, 3352
(1994),  \  and Phys.\ Rev.\ \textbf{D50}, 2225 (1994).


\bibitem{MunPesch}
S.~Munier and R.~B.~Peschanski,
%``Geometric scaling as traveling waves,''
Phys.\ Rev.\ Lett.\ \textbf{91} (2003) 232001 [hep-ph/0309177], and
%%CITATION = HEP-PH/0309177;%%
%``Traveling wave fronts and the transition to saturation,''
Phys.\ Rev.\ D \textbf{69} (2004) 034008 [hep-ph/0310357].
%%CITATION = HEP-PH/0310357;%%
\bibitem{Forte}
 %\cite{Caola:2008xr}
%\bibitem{Caola:2008xr} 
  F.~Caola and S.~Forte,
%\emph{Geometric Scaling from GLAP evolution},
{Phys.\ Rev.\ Lett.}  {\bf 101} (2008) 022001  
  [arXiv:0802.1878 [hep-ph]].
  %%CITATION = ARXIV:0802.1878;%% 



\end{thebibliography}
\end{document}